\documentclass[preprint,12pt]{elsarticle}
\usepackage[pdfborder={0 0 0}]{hyperref} 
\usepackage{amsthm,amsmath,amsfonts,latexsym,amssymb} 
\usepackage{graphics,fancyhdr,graphicx,subfigure} 
\usepackage[ruled,linesnumbered]{algorithm2e} 
\usepackage[margin=1in]{geometry}
\usepackage{multirow,booktabs} 
\usepackage{appendix}
\usepackage[table]{xcolor}
\usepackage{listings}
\usepackage{tabularx}
\usepackage{placeins}
\usepackage{comment}
\usepackage{lineno}
\usepackage{lscape}
\usepackage{bm}
\usepackage{hyperref}
\usepackage{mathrsfs}
\usepackage{arydshln}
\usepackage{appendix}
\usepackage{float} 
\usepackage{enumitem}
\usepackage{url}
\usepackage{caption}
\usepackage{nicematrix}
\usepackage{makecell} 
\usepackage[
singlelinecheck=false 
]{caption} 
\usepackage{caption}
\newcolumntype{C}[1]{>{\centering\arraybackslash}p{#1}}
\usepackage{upgreek} 

\definecolor{custom-blue}{RGB}{3,69,173}
\hypersetup{
    colorlinks = true,
    urlcolor   = blue,
    citecolor  = custom-blue,
}

\biboptions{numbers,sort&compress} 
\definecolor{listinggray}{gray}{0.9}
\definecolor{lbcolor}{rgb}{0.9,0.9,0.9}
\definecolor{Darkgreen}{RGB}{0,100,0}

\newcommand{\WW}[2][black]{{\textcolor{#1}{#2}}}
\begin{document} 
\abovedisplayskip=6.0pt
\belowdisplayskip=6.0pt
\begin{frontmatter} 

\title{A Non-Overlapping Schwarz Hybrid Finite Element–Neural Operator Framework for Solid Mechanics on Irregular Domains}

\author[1,2]{Wei Wang}
\ead{wwang198@jhu.edu}
\author[3]{Abhinav Gupta}
\ead{abhinav.gupta@vanderbilt.edu}
\author[1,4]{Haihui Ruan \corref{cor1}}
\ead{hhruan@polyu.edu.hk}
\author[2]{Somdatta Goswami\corref{cor1}}
\ead{somdatta@jhu.edu}

\address[1]{Department of Mechanical Engineering, The Hong Kong Polytechnic University}
\address[2]{Department of Civil and Systems Engineering, Johns Hopkins University}
\address[3]{Civil and Environmental Engineering, Vanderbilt University}
\address[4]{PolyU-Daya Bay Technology and Innovation Research Institute}
\cortext[cor1]{Corresponding author.}

\begin{abstract}
Finite element (FE) methods are the benchmark for solid mechanics simulations, yet their computational cost becomes prohibitive in problems exhibiting localised nonlinearities, fine-scale features, or long-time dynamic evolution. In our earlier FE-neural operator (FE-NO) hybrid framework~\cite{wang2025time}, physics-informed deep operator networks were coupled with FE solvers through overlapping domain decomposition with Dirichlet-Dirichlet interface exchange, accelerating computationally intensive subdomains employing NOs while preserving FE fidelity elsewhere. Two limitations remained: the overlapping formulation required redundant interface computations that increased inner Schwarz iteration counts, and the convolutional feature extractor restricted the NO subdomain to structured grids, precluding irregular geometries. This work addresses both limitations. A non-overlapping Schwarz alternating method with Neumann-Dirichlet interface exchange replaces the overlapping formulation, transmitting traction boundary conditions from the NO to FE rather than exchanging displacement. This eliminates the overlap layer and reduces the number of inner Schwarz iterations, while maintaining bounded error accumulation across all tested time horizons. To enable simulations on arbitrarily shaped subdomains, a Point-DeepONet is introduced, which directly operates on unstructured FE point clouds without interpolation, extending the framework to non-convex and irregular subdomain geometries. Furthermore, strain and stress operators are derived analytically from the displacement operators through the governing kinematic equations, rather than trained as independent networks, significantly reducing the number of trainable parameter sets, while enforcing mechanical consistency by construction. The framework is validated on three solid mechanics benchmarks: static linear elasticity, quasi-static hyperelasticity, and elastodynamics with both regular and irregular subdomain geometries. These results establish a non-overlapping FE-NO coupling paradigm that is geometry-flexible, parameter-efficient, and convergence-stable, providing a principled pathway for hybrid physics-based and operator-learning solvers in large-scale dynamic solid mechanics. 

\end{abstract}
\end{frontmatter}

\section{Introduction}
\label{sec:intro}

Partial differential equations (PDEs) remain central to computational solid mechanics, yet achieving high-fidelity simulations often requires fine discretizations and repeated nonlinear solves. These challenges become particularly severe in multiscale and dynamic systems, where localized fine-scale features and long-time temporal evolution substantially increase computational cost. Recent advances in scientific machine learning (SciML) have demonstrated that Neural Operators (NOs)~\cite{rabczuk2026scientific, lu2021learning,li2021fourier, tripura2023wavelet, raonic2023convolutional, cao2024laplace} provide an efficient alternative for approximating solution operators of PDE-governed systems~\cite{pestourie2023physics, degen2023perspectives, brunton2024promising, koh2025recent, agarwal2026multimodal, goswami2023physics, li2025importance}. Once trained, NOs can rapidly predict solutions for a family of PDEs at a fraction of the computational cost of traditional numerical solvers~\cite{azizzadenesheli2024neural,kota2026hybrid}. 

Despite these advantages, standalone NOs still face several limitations that restrict their deployment in large-scale engineering simulations. First, many NO architectures require large volumes of high-fidelity training data generated from expensive numerical solvers. Second, prediction accuracy and computational efficiency often deteriorate for large computational domains containing localized nonlinearities or fine-scale physical phenomena. Physics-informed operator learning approaches, such as Physics-Informed Deep Operator Networks (PI-DeepONet)~\cite{goswami2023physics, garg2026spinonet, mandl2025separable, sarkar2026learning}, alleviate the first challenge by embedding governing equations directly into the training process, thereby significantly reducing data requirements. However, the second limitation remains largely unresolved, particularly for complex multiscale dynamical systems in which accuracy, stability, and scalability must be maintained simultaneously.

To address the second limitation, a growing body of work has coupled NOs with established numerical solvers to form hybrid computational frameworks \cite{amin2026fenn, pantidis2025integrated, actor2023machine,mota2025fundamentally,chung2026latent,ouyang2026noem,puthli2026neural,roy2025best,mandl2026physics}. Domain decomposition methods (DDMs) provide the natural coupling mechanism: computationally intensive subdomains are delegated to pretrained neural surrogates, while conventional FE solvers govern the remainder where coarse discretizations remain adequate. Within this paradigm, the choice of interface coupling, overlapping vs.\ non-overlapping, Dirichlet-Dirichlet vs.\ Neumann-Dirichlet, critically determines convergence rate, interface communication cost, and geometric flexibility of the neural surrogate. In our previous FE-NO framework~\cite{wang2025time}, PI-DeepONet was coupled with FE solvers through an overlapping Schwarz alternating method with Dirichlet-Dirichlet interface exchange, demonstrating that neural surrogates can substantially accelerate selected subdomains while preserving global solution accuracy. That framework also introduced a time-marching DeepONet architecture by embedding the Newmark-$\beta$ scheme into the neural operator, enabling stable long-time dynamic prediction. Two limitations remained: the overlapping formulation required redundant interface computations that increased inner Schwarz iteration counts, and the convolutional branch network restricted the NO subdomain to structured spatial grids, precluding irregular or unstructured geometries. The present work addresses both limitations directly.

\begin{figure}[H]
\begin{center}
\includegraphics[width=1\textwidth]{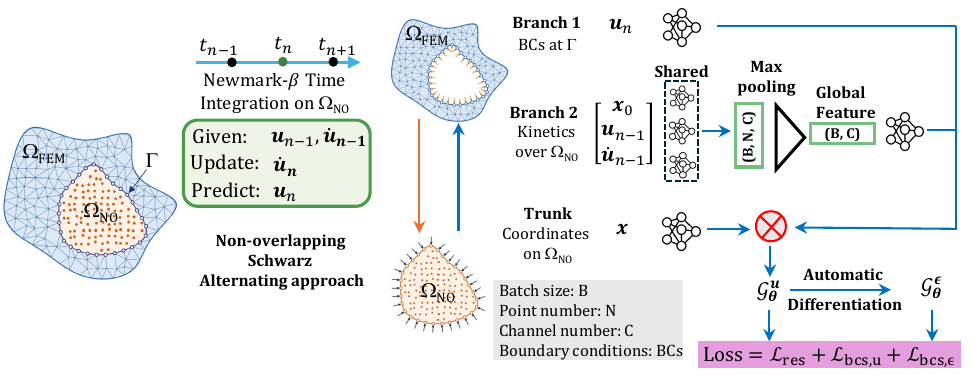}
\caption{Schematic of the proposed framework: Non-overlapping Schwarz alternating approach built on a time-marching Point-DeepONet. The computational domain is partitioned into two non-overlapping regions,$\Omega_{\mathrm{FEM}}$ and $\Omega_{\mathrm{NO}}$, solved by FE solver and a PI-DeepONet, respectively, and coupled through their shared boundary $\Gamma$, across which the boundary information is exchanged at each Schwarz iteration. Within time-marching Point-DeepONet, the solution is advanced by the Newmark-$\beta$ scheme: given the kinetic state $(\mathbf{u}_{n-1}, \dot{\mathbf{u}}_{n-1})$, the network predicts the $\mathbf{u}$ at current time step $n$, then updates the velocity $\dot{\mathbf{u}}_n$. Branch 1 (an MLP) encodes Dirichlet boundary conditions $u_{|\Gamma}$ at current time step $n$, while Branch 2 (PointNet + MLP) encodes the kinetic state $(\mathbf{u},\dot{\mathbf{u}})_{|\Omega_{\mathrm{NO}}}$ at time step $n-1$ together with its reference coordinates $\mathbf{x}_0$. The two branches produce embedding with identical dimension, whose dot product with the trunk output - evaluated at the spatial coordinates $\mathbf{x}$ - yields the displacement operators $\mathcal{G}^{u_x}_{\bm\theta_1}$ and $\mathcal{G}^{u_y}_{\bm\theta_2}$, from which strain operators are derived analytically via Eqs.~\eqref{eq:eps_xx}-\eqref{eq:eps_yy}. }
\label{Fig:DeepONet_structure}
\end{center}
\end{figure}

In the present work, we extend the previously developed FE-NO framework through a non-overlapping domain decomposition strategy combined with a PointNet-enhanced time-marching neural operator architecture. The proposed framework (see \WW{Figure \ref{Fig:DeepONet_structure}}) couples FE solvers and PI-DeepONets through a non-overlapping Schwarz alternating method~\cite{lions1988schwarz,haase1992non}, where Neumann-Dirichlet information exchange is enforced at subdomain interfaces. Compared to overlapping formulations, the proposed approach reduces redundant interface solves and improves convergence efficiency while maintaining robust spatiotemporal coupling. Within this framework, PI-DeepONet is assigned to computationally intensive subdomains characterized by localized nonlinearities or fine-scale features, whereas conventional FE solvers govern the remainder of the computational domain. To address dynamic simulations on arbitrarily shaped subdomains, we further introduce a time-marching Point-DeepONet by integrating PointNet~\cite{qi2017pointnet, park2024point} into the branch network of DeepONet. Unlike previous grid-dependent implementations, the proposed architecture directly processes unstructured point-cloud representations without interpolation, extending the framework toward complex geometries and future three-dimensional applications. In addition, the Newmark time-stepping scheme~\cite{newmark1959method} is embedded directly into the neural operator architecture, enabling stable temporal coupling between the FE and NO subdomains while mitigating long-term error accumulation in autoregressive predictions. The main contributions of this work are summarized as follows:
\begin{itemize}[leftmargin=*,nosep]
    \item The stress and strain operators are derived analytically from the displacement operators via kinematic equations, rather than trained independently, reducing the number of free parameters and enforcing physical consistency.
    
    \item PointNet is embedded in the time-marching branch network, enabling interpolation-free extraction of kinetic state from unstructured FE meshes and allowing the NO subdomain to take arbitrary shapes - a prerequisite for three-dimensional applications. 
    
    \item The coupling interface is reformulated from an overlapping Dirichlet-Dirichlet to a non-overlapping Neumann-Dirichlet approach, which is shown to reduce the number of inner Schwarz iterations from 9 to 3 per time step in elasto-dynamic simulations while maintaining bounded error accumulation.
\end{itemize}

\section{Methodology}
\label{sec:methodology}
 
Building on the overlapping Dirichlet–Dirichlet FE-NO coupling framework established in~\cite{wang2025time}, this work proposes a non-overlapping Neumann–Dirichlet hybrid solver that advances the robustness and computational efficiency of domain decomposition methods for solid mechanics. Three loading regimes are addressed: static linear elasticity, quasi-static hyperelasticity, and elastodynamics. In all cases, the momentum balance over domain $\Omega$ reads:
\begin{alignat}{2}
  &\text{Static / quasi-static:} \quad & \nabla \cdot \bm{\sigma} + \mathbf{f} &= \mathbf{0}, \label{eq:static} \\
  &\text{Dynamic:}               \quad & \nabla \cdot \bm{\sigma} + \mathbf{f} &= \rho\ddot{\mathbf{u}}, \label{eq:dynamic}
\end{alignat}
where $\bm{\sigma}$, $\mathbf{f}$, $\mathbf{u}$, and $\rho$ denote the Cauchy stress tensor, body force, displacement vector, and mass density, respectively.
 
PI-DeepONet is implemented in JAX~\cite{jax2018github}; all FE models are built in FEniCSx~\cite{baratta2023dolfinx} with triangular meshes generated by Gmsh~\cite{geuzaine2009gmsh}. A second-order continuous Galerkin discretization is used throughout to ensure stress and strain continuity across the FE subdomain. The DeepONet parameterizes a nonlinear operator $\mathcal{G}:\mathcal{U} \to \mathcal{S}$ between Banach spaces of input, $\mathcal{U}$, and output, $\mathcal{S}$, functions, where
\begin{equation}
  \mathcal{U} = \bigl\{\Omega^{*};\; h : X \to \mathbb{R}^{d_x}\bigr\},
  \quad X \subseteq \mathbb{R}^{d_h},
  \label{eq:U}
\end{equation}
\begin{equation}
  \mathcal{S} = \bigl\{\Omega^{*};\; s : Y \to \mathbb{R}^{d_y}\bigr\},
  \quad Y \subseteq \mathbb{R}^{d_s}.
  \label{eq:S}
\end{equation}
Here, $\Omega^{*}$ denotes the DeepONet subdomain; $h$ and $s$ are the input and output functions, respectively; $X \subseteq \mathbb{R}^{d_h}$ and $Y \subseteq \mathbb{R}^{d_s}$ are the corresponding domains of definition, with $d_h$ and $d_s$ their respective dimensions; and $d_x$ and $d_y$ denote the output dimensionalities of $h$ and $s$. Following~\cite{lu2021learning}, the parameterised form is:
\begin{equation}
  \mathcal{G}_{\bm{\theta}} : \mathcal{U} \to \mathcal{S},
  \qquad \bm{\theta} \in \bm\Theta,
  \label{eq:deeponet}
\end{equation}
where $\bm\Theta$ is a finite-dimensional parameter space and $\bm{\theta}$ collects all trainable weights and biases. Two distinct PI-DeepONet architectures are developed for the static/quasi-static and dynamic regimes; their network dimensions are given in Table~\ref{tab:arch}. In both cases, optimal parameters $\bm{\theta}^{*}$ are obtained by minimizing a composite loss function assembled from the governing PDE residuals and boundary conditions, as detailed in Sections~\ref{sec:arch} and~\ref{sec:ddm}.
 
\subsection{Governing Equations}
\label{sec:governing}
 
\paragraph{Linear elasticity (static and dynamic)}
Under the small-deformation assumption, strain and stress are related to displacement by
\begin{equation}
  \bm{\varepsilon} = \tfrac{1}{2}\bigl[\nabla\mathbf{u}
                     + (\nabla\mathbf{u})^{\mathsf{T}}\bigr],
  \label{eq:strain_linear}
\end{equation}
\begin{equation}
  \bm{\sigma} = \lambda\,\mathrm{tr}(\bm{\varepsilon})\mathbf{I}
                + 2\mu\bm{\varepsilon},
  \label{eq:stress_linear}
\end{equation}
where $\lambda$ and $\mu$ are the Lamé constants. Substituting into Eq.~\eqref{eq:static} yields a linear elliptic problem discretized by FE as
\begin{equation}
  \mathbf{K}\mathbf{u} = \mathbf{F},
  \label{eq:FE_static}
\end{equation}
with global stiffness matrix $\mathbf{K}$ and external force vector $\mathbf{F}$. Both $\mathbf{u}$ and $\bm{\varepsilon}$ satisfy the same governing equation, which is exploited directly in the PI-DeepONet loss formulation (Section~\ref{sec:arch}). The dynamic counterpart is treated below.
 
\paragraph{Hyperelasticity (quasi-static)}
Under large deformations, the kinematic state is described by the deformation gradient $\mathbf{F}_{g} = \mathbf{I} + \nabla\mathbf{u}$, the right Cauchy-Green tensor $\mathbf{C} = \mathbf{F}_{g}^{\mathsf{T}} \mathbf{F}_{g}$, its first invariant $I_1 = \mathrm{tr}(\mathbf{C})$, and the Jacobian $J = \det\mathbf{F}_{g}$. The Green-Lagrange strain is
\begin{equation}
  \mathbf{E}_{\mathrm{GL}} = \tfrac{1}{2}(\mathbf{C} - \mathbf{I}),
  \label{eq:EGL}
\end{equation}
and the second Piola-Kirchhoff stress reads
\begin{equation}
  \mathbf{S} = \mu\bigl(\mathbf{I} - \mathbf{C}^{-1}\bigr)
               + \lambda(\ln J)\mathbf{C}^{-1}.
  \label{eq:PK2}
\end{equation}

The Cauchy stress is recovered as
\begin{equation}
  \bm{\sigma} = \frac{\mathbf{F}_{g}\,\mathbf{S}\,\mathbf{F}_{g}^{\mathsf{T}}}{J}.
  \label{eq:Cauchy_large}
\end{equation}
A Neo-Hookean strain energy density,
\begin{equation}
  \Psi = \tfrac{1}{2}\mu(I_1 - 3) + \tfrac{1}{2}\lambda(\ln J)^{2}
         - \mu\ln J,
  \label{eq:neo_hookean}
\end{equation}
yields the first Piola-Kirchhoff stress $\mathbf{P} = \partial\Psi / \partial\mathbf{F}_{g} = \mathbf{F}_{g}\mathbf{S}$. In the absence of body forces, quasi-static equilibrium at time step $n$ requires
\begin{equation}
  \nabla \cdot \mathbf{P}^{n} = \mathbf{0}.
  \label{eq:qs_equil}
\end{equation}
The strong nonlinearity of Eqs.~\eqref{eq:EGL}-\eqref{eq:qs_equil} necessitates Newton-Raphson iteration (Newton's solver in PETSc~\cite{petsc-web-page}).
 
\paragraph{elastodynamics} Neglecting damping and discretizing only the spatial dimensions, Eq.~\eqref{eq:dynamic} becomes
\begin{equation}
  \mathbf{M}\ddot{\mathbf{u}}^{n} + \mathbf{K}\mathbf{u}^{n}
  = \mathbf{F}^{n},
  \label{eq:semi_discrete}
\end{equation}
where $\mathbf{M}$ is the global mass matrix and the superscript $n$
denotes the current time step. Time integration follows the Newmark-$\beta$
scheme~\cite{newmark1959method}, with $\gamma = \beta = \tfrac{1}{2}$ chosen for
unconditional stability, giving the update formulas
\begin{align}
  \dot{\mathbf{u}}^{n} &= \dot{\mathbf{u}}^{n-1}
    + \tfrac{\Delta t}{2}\bigl(\ddot{\mathbf{u}}^{n-1}
    + \ddot{\mathbf{u}}^{n}\bigr),
  \label{eq:newmark_v} \\
  \mathbf{u}^{n} &= \mathbf{u}^{n-1}
    + \Delta t\,\dot{\mathbf{u}}^{n-1}
    + \tfrac{\Delta t^{2}}{2}\ddot{\mathbf{u}}^{n}.
  \label{eq:newmark_u}
\end{align}
Substituting Eq.~\eqref{eq:newmark_u} into Eq.~\eqref{eq:semi_discrete} and assuming zero body force ($\mathbf{F}^{n} = \mathbf{0}$) yields the effective system
\begin{equation}
  \left(\frac{2}{\Delta t^{2}}\mathbf{M} + \mathbf{K}\right)\mathbf{u}^{n}
  = \frac{2}{\Delta t^{2}}\mathbf{M}\mathbf{u}^{n-1}
    + \frac{2}{\Delta t}\mathbf{M}\dot{\mathbf{u}}^{n-1},
  \label{eq:effective_system}
\end{equation}
which has the same algebraic structure as Eq.~\eqref{eq:FE_static} and is therefore assembled and solved identically within the FE framework.

\subsection{PI-DeepONet Architecture and Training}
\label{sec:arch}
 
\paragraph{Network structure}
The standard DeepONet~\cite{lu2021learning} comprises a branch network encoding
input function information and a trunk network encoding the spatiotemporal query coordinates;
their outputs are combined via a dot product to form the solution operator.
\WW{Two branch--trunk networks with identical architecture but separate (non-shared) parameter sets $\bm\theta_1$, and $\bm\theta_2$ are trained jointly, each predicting one displacement component, yielding the operators $\mathcal{G}^{u_x}_{\bm\theta_1}$ and $\mathcal{G}^{u_y}_{\bm\theta_2}$ for the $x$- and $y$-components, respectively. }Under static or quasi-static loading, Branch~1 encodes only
the Dirichlet boundary data $\mathbf{u}|_{\partial\Omega}$; Branch 2 net is
activated exclusively for the dynamic surrogate (see below Figure \ref{Fig:DeepONet_structure}). 
 
\paragraph{Analytically Derived Strain and Stress Operators}
Rather than parameterizing the strain components independently through DeepONets, all three strain operators are derived analytically from the displacement operators
through the kinematic relation~\eqref{eq:strain_linear}:
\begin{align}
  \mathcal{G}^{\varepsilon_{xx}}_{\bm\theta_1,\bm\theta_2}
    &= \frac{\partial \mathcal{G}^{u_x}_{\bm\theta_1}}{\partial x},
    \label{eq:eps_xx} \\
  \mathcal{G}^{\varepsilon_{xy}}_{\bm\theta_1,\bm\theta_2}
    &= \frac{1}{2}\!\left(
       \frac{\partial \mathcal{G}^{u_x}_{\bm\theta_1}}{\partial y}
       + \frac{\partial \mathcal{G}^{u_y}_{\bm\theta_2}}{\partial x}
       \right),
    \label{eq:eps_xy} \\
  \mathcal{G}^{\varepsilon_{yy}}_{\bm\theta_1,\bm\theta_2}
    &= \frac{\partial \mathcal{G}^{u_y}_{\bm\theta_2}}{\partial y}.
    \label{eq:eps_yy}
\end{align}
Consequently, all five field operators share only the two parameter sets
$\{\bm\theta_1, \bm\theta_2\}$, substantially reducing GPU memory and compute
costs while enforcing kinematic consistency by construction. For the
hyperelastic case, the Cauchy stress components $\sigma_{xx}$,
$\sigma_{xy}$, $\sigma_{yy}$ replace the strain operators and are
similarly expressed as closed-form functions of
$\mathcal{G}^{u_x}_{\bm\theta_1}$ and $\mathcal{G}^{u_y}_{\bm\theta_2}$ via
Eqs.~\eqref{eq:EGL}-\eqref{eq:Cauchy_large}; the two-parameter structure
is retained.
 
\paragraph{PointNet branch for dynamics}
In the elasto-dynamic case, the domain kinetic state at the previous time
step - namely $(u_x^{n-1}, u_y^{n-1}, \dot{u}_x^{n-1},
\dot{u}_y^{n-1})$ together with their associated spatial coordinates -
must be ingested by Branch net. Compared to our prior work~\cite{wang2025time},
which used a convolutional neural network (CNN) and therefore required
regular grids, we replace Branch net with a simplified PointNet~\cite{park2024point}, allowing it to operate on unstructured FE meshes of arbitrary shape without interpolation.
 
The PointNet accepts point clouds of shape $(B,\,N,\,6)$, where $B$ is
the batch size, $N$ the number of points, and the six channels concatenate
$(x_0, y_0, u_x^{n-1}, u_y^{n-1}, \dot{u}_x^{n-1}, \dot{u}_y^{n-1})$.
Three one-dimensional convolutional layers, each followed by a $\tanh$
activation, act as a shared MLP applied point-wise to extract localised
geometric and kinetic features - mirroring the local-interaction
principle of FE, in which only neighbouring nodes substantially influence
computations at a given node. A global max-pooling layer then aggregates
per-point embeddings into a fixed-length representation, which is passed
through a fully connected network to produce the Branch net output. Because
the architecture is invariant to point ordering and mesh topology, it
extends naturally to non-convex and three-dimensional domains.
 
\paragraph{Composite loss function.}
The PI-DeepONet training is formulated as a boundary value problem over
$\Omega$:
\begin{align}
  \mathcal{N}\bigl[\mathcal{G}_{\bm{\theta}}(\mathbf{h})\bigr](\mathbf{x})
    &= \mathbf{0}, \quad \mathbf{x} \in \Omega, \label{eq:pde_loss} \\
  \mathcal{B}\bigl[\mathcal{G}_{\bm{\theta}}(\mathbf{h})\bigr](\mathbf{x})
    &= \mathbf{g}(\mathbf{x}), \quad \mathbf{x} \in \partial\Omega,
    \label{eq:bc_loss}
\end{align}
where $\mathcal{N}$ and $\mathcal{B}$ are the PDE and boundary
differential operators, and $\mathbf{g}$ is the prescribed boundary data.
No body force is considered, all boundary conditions are of Dirichlet
type, and a plane-strain assumption is adopted.
 
The complete set of five operators is denoted
\begin{equation}
  \mathcal{G}(\mathbf{h},\bm{\theta})
  = \bigl\{
      \mathcal{G}^{u_x}_{\bm\theta_1},\;
      \mathcal{G}^{u_y}_{\bm\theta_2},\;
      \mathcal{G}^{\varepsilon_{xx}}_{\bm\theta_1,\bm\theta_2},\;
      \mathcal{G}^{\varepsilon_{xy}}_{\bm\theta_1,\bm\theta_2},\;
      \mathcal{G}^{\varepsilon_{yy}}_{\bm\theta_1,\bm\theta_2}
    \bigr\},
  \label{eq:operator_set}
\end{equation}
with input $\mathbf{h} = (u_x, u_y, \varepsilon_{xx}, \varepsilon_{xy}, \varepsilon_{yy})$. Substituting Eq.~\eqref{eq:operator_set} into
Eqs.~\eqref{eq:pde_loss}-\eqref{eq:bc_loss} and collecting residuals at
$N^r$ interior collocation points $\{{\mathbf{x}^r_i}\}_{i=1}^{N^r}
\subset \Omega$ and $N^b$ boundary points
$\{{\mathbf{x}^b_i}\}_{i=1}^{N^b} \subset \partial\Omega$, the composite
loss function is
\begin{equation}
\begin{split}
  \mathcal{L}(\bm{\theta})
  &= \mathcal{L}_{\mathrm{res}}(\bm{\theta})
   + \mathcal{L}_{\mathrm{bcs},u}(\bm{\theta})
   + \mathcal{L}_{\mathrm{bcs},\varepsilon}(\bm{\theta}) \\
  &= \frac{1}{N^r}\sum_{i=1}^{N^r}
       \left\|\mathcal{N}_1\!\left(
         \mathcal{G}^{u_x}_{\bm\theta_1},\mathcal{G}^{u_y}_{\bm\theta_2}
       \right)(\mathbf{x}^r_i)\right\|^2
   + \frac{1}{N^r}\sum_{i=1}^{N^r}
       \left\|\mathcal{N}_2\!\left(
         \mathcal{G}^{u_x}_{\bm\theta_1},\mathcal{G}^{u_y}_{\bm\theta_2}
       \right)(\mathbf{x}^r_i)\right\|^2 \\
  &\quad
   + \frac{1}{N^b}\sum_{i=1}^{N^b}
       \left\|\mathcal{G}^{u_x}_{\bm\theta_1}(\mathbf{x}^b_i)
              - s_1(\mathbf{x}^b_i)\right\|^2
   + \frac{1}{N^b}\sum_{i=1}^{N^b}
       \left\|\mathcal{G}^{u_y}_{\bm\theta_2}(\mathbf{x}^b_i)
              - s_2(\mathbf{x}^b_i)\right\|^2 \\
  &\quad
   + \frac{1}{N^b}\sum_{i=1}^{N^b}
       \left\|\mathcal{G}^{\varepsilon_{xx}}_{\bm\theta_1,\bm\theta_2}(\mathbf{x}^b_i)
              - s_3(\mathbf{x}^b_i)\right\|^2
   + \frac{1}{N^b}\sum_{i=1}^{N^b}
       \left\|\mathcal{G}^{\varepsilon_{xy}}_{\bm\theta_1,\bm\theta_2}(\mathbf{x}^b_i)
              - s_4(\mathbf{x}^b_i)\right\|^2 \\
  &\quad
   + \frac{1}{N^b}\sum_{i=1}^{N^b}
       \left\|\mathcal{G}^{\varepsilon_{yy}}_{\bm\theta_1,\bm\theta_2}(\mathbf{x}^b_i)
              - s_5(\mathbf{x}^b_i)\right\|^2,
\end{split}
\label{eq:loss}
\end{equation}
where $\mathcal{N}_1$ and $\mathcal{N}_2$ are the $x$- and
$y$-direction equilibrium residual operators, and $\mathbf{s} =
(s_1,\ldots,s_5)$ is the target output function in $\mathcal{S}$.
Minimizing $\mathcal{L}$ yields the optimal parameters
\begin{equation}
  \bm{\theta}^{*} = \{\bm\theta_1^{*},\,\bm\theta_2^{*}\}
  = \arg\min_{\bm{\theta}}\,\mathcal{L}(\bm{\theta}).
  \label{eq:opt}
\end{equation}
 
\paragraph{Training data} Boundary conditions $\mathbf{u}|_{\partial\Omega}$,
strain (or stress) boundary values, and prior-step kinetic fields
$\mathbf{u}^{n-1}|_{\Omega}$, $\dot{\mathbf{u}}^{n-1}|_{\Omega}$ are
sampled from Gaussian Random Fields (GRF) to maximize
generalization ability (see Appendix~A). \WW{Note: only the boundary strains are labeled with corresponding displacement loads as input-output pairs of training data generated for the FE solver; the NOs are trained in physics-informed manner only on the collocation points in terms of governing PDEs.}
 
Once pretrained, the displacement operators
$\mathcal{G}^{u_x}_{\bm\theta_1^{*}}$ and
$\mathcal{G}^{u_y}_{\bm\theta_2^{*}}$ predict $\mathbf{u}^{n}|_{\Omega}$
given the boundary conditions at time step $n$ and the kinetic state from
step $n{-}1$; velocity $\dot{\mathbf{u}}^{n}$ is then updated via
Eq.~\eqref{eq:newmark_v}. At each step, the strain operators furnish
traction along the non-overlapping interface, which serves as the Neumann
boundary condition for the FE solver (Section~\ref{sec:ddm}).
 
For hyperelasticity, the Cauchy stress components $(\sigma_{xx},
\sigma_{xy}, \sigma_{yy})$ replace the strain operators in
Eq.~\eqref{eq:loss}, expressed through
$\mathcal{G}^{u_x}_{\bm\theta_1}$ and $\mathcal{G}^{u_y}_{\bm\theta_2}$ via
Eqs.~\eqref{eq:PK2}-\eqref{eq:Cauchy_large}. The network structure,
loss formulation, and coupling workflow are otherwise identical to the
linear-elastic case.

\subsection{Non-Overlapping Domain Decomposition}
\label{sec:ddm}

Spatial coupling between the FE and NO solvers is achieved via a Schwarz alternating method~\cite{mota2017schwarz} on a non-overlapping partition
$\Omega = \Omega_{\mathrm{I}} \cup \Omega_{\mathrm{II}}$ with
$\Omega_{\mathrm{I}} \cap \Omega_{\mathrm{II}} = \emptyset$
(Figure~\ref{Fig:DeepONet_structure}). The FE solver governs $\Omega_{\mathrm{I}}$; the
pretrained NO governs $\Omega_{\mathrm{II}}$. Their shared interface
$\Gamma_{\mathrm{II}} = \Gamma^{\mathrm{in}}_{\mathrm{I}}$ - noted as $\Gamma$ in Figure \ref{Fig:DeepONet_structure} - is the sole
boundary of $\Omega_{\mathrm{II}}$ and the inner boundary of
$\Omega_{\mathrm{I}}$; the outer boundary $\Gamma^{\mathrm{out}}_{\mathrm{I}}$
carries the external loading.
 
Compared to the overlapping Dirichlet-Dirichlet scheme of~\cite{wang2025time}, the Neumann-Dirichlet approach adopted here transmits interface traction from the NO to FE rather than exchanging displacements between two FE
solvers. This eliminates the overlap layer and reduces the number of inner
Schwarz iterations per time step from approximately ten (as observed
in~\cite{wang2025time}) to three in the elasto-dynamic case, while
maintaining bounded error accumulation across all time steps.
 
The coupling procedure at each time step $n$ is formalised in
Algorithm~\ref{algo_1}. Starting from $\mathbf{u}^{n,0} = \mathbf{0}$
in both subdomains, each inner iteration $j$ proceeds as follows: (i) FE
solves $\Omega_{\mathrm{I}}$ given the interface traction
$\mathbf{T}^{n,j-1}|_{\Gamma^{\mathrm{in}}_{\mathrm{I}}}$ from the
previous NO solve and the external Dirichlet data on
$\Gamma^{\mathrm{out}}_{\mathrm{I}}$; (ii) the interface displacement
$\mathbf{u}^{n,j}|_{\Gamma^{\mathrm{in}}_{\mathrm{I}}}$ is passed to the
NO as a Dirichlet condition; (iii) the NO solves $\Omega_{\mathrm{II}}$
and returns $\bm{\sigma}^{n,j}|_{\Gamma_{\mathrm{II}}}$; (iv) a relaxed
traction is computed as
\begin{equation}
  \tilde{\bm{\sigma}}^{n,j}|_{\Gamma_{\mathrm{II}}}
  = (1 - \rho_r)\,\bm{\sigma}^{n,j}|_{\Gamma_{\mathrm{II}}}
  + \rho_r\,\bm{\sigma}^{n,j}|_{\Gamma^{\mathrm{in}}_{\mathrm{I}}},
  \label{eq:relaxation}
\end{equation}
with relaxation parameter $\rho_r = 0.5$, and the interface traction is updated as
$\mathbf{T}^{n,j}|_{\Gamma^{\mathrm{in}}_{\mathrm{I}}}
= \tilde{\bm{\sigma}}^{n,j}|_{\Gamma_{\mathrm{II}}}
\cdot \mathbf{n}|_{\Gamma^{\mathrm{in}}_{\mathrm{I}}}$.
 
The inner iteration terminates when the $L^2$-norm of successive
displacement increments falls below a threshold $\epsilon$:
\begin{equation}
  \bigl\|\mathbf{u}^{n,j}|_{\Omega_{\mathrm{I}}}
         - \mathbf{u}^{n,j-1}|_{\Omega_{\mathrm{I}}}\bigr\|_{L^2}
  + \bigl\|\mathbf{u}^{n,j}|_{\Omega_{\mathrm{II}}}
           - \mathbf{u}^{n,j-1}|_{\Omega_{\mathrm{II}}}\bigr\|_{L^2}
  < \epsilon.
  \label{eq:convergence}
\end{equation}
Upon convergence, subdomain fields are assembled as
\begin{alignat}{2}
  \mathbf{u}^{n}|_{\Omega_{\mathrm{I}}}   &:= \mathbf{u}^{n,j}|_{\Omega_{\mathrm{I}}},   &\qquad
  \mathbf{u}^{n}|_{\Omega_{\mathrm{II}}}  &:= \mathbf{u}^{n,j}|_{\Omega_{\mathrm{II}}},  \label{eq:assemble_u} \\
  \bm{\varepsilon}^{n}|_{\Omega_{\mathrm{I}}}   &:= \bm{\varepsilon}^{n,j}|_{\Omega_{\mathrm{I}}},   &\qquad
  \bm{\varepsilon}^{n}|_{\Omega_{\mathrm{II}}}  &:= \bm{\varepsilon}^{n,j}|_{\Omega_{\mathrm{II}}},  \label{eq:assemble_eps} \\
  \bm{\sigma}^{n}|_{\Omega_{\mathrm{I}}}   &:= \bm{\sigma}^{n,j}|_{\Omega_{\mathrm{I}}},   &\qquad
  \bm{\sigma}^{n}|_{\Omega_{\mathrm{II}}}  &:= \bm{\sigma}^{n,j}|_{\Omega_{\mathrm{II}}},  \label{eq:assemble_sig}
\end{alignat}
yielding the global fields $\mathbf{u}|_{\Omega}$,
$\bm{\varepsilon}|_{\Omega}$, and $\bm{\sigma}|_{\Omega}$ by merging the
two subdomains. For the static regime, the outer time loop is omitted and
the inner iteration operates identically.
 
\paragraph{Interface normal under large deformation.}
For small deformations, the outward unit normal to the circular interface
$\Gamma^{\mathrm{in}}_{\mathrm{I}}$ is
\begin{equation}
  \mathbf{n}|_{\Gamma^{\mathrm{in}}_{\mathrm{I}}}
  = -\left(\frac{x - x_c}{r_0},\;\frac{y - y_c}{r_0}\right),
  \label{eq:normal_small}
\end{equation}
where $(x_c, y_c)$ is the circle centre and $r_0$ its radius. Under large
deformations, reference and current surface normals are related by
Nanson's formula~\cite{destrade2025deformations}:
\begin{equation}
  \mathbf{n}\,\mathrm{d}A
  = J\mathbf{F}_{g}^{-\mathsf{T}}\,\mathbf{n}_{R}\,\mathrm{d}A_{R},
  \label{eq:nanson}
\end{equation}
where $\mathbf{n}_{R}$ is the reference normal given by
Eq.~\eqref{eq:normal_small}. The current outward unit normal is therefore
\begin{equation}
  \mathbf{n}|_{\Gamma^{\mathrm{in}}_{\mathrm{I}}}
  = \frac{J\mathbf{F}_{g}^{-\mathsf{T}}\,\mathbf{n}_{R}}
         {\left\|J\mathbf{F}_{g}^{-\mathsf{T}}\,\mathbf{n}_{R}\right\|},
  \label{eq:normal_large}
\end{equation}
which enters the weak form of the FE traction boundary condition in the hyperelastic case. Note that Algorithm~\ref{algo_1} is equally applicable to an FE-FE coupling scheme (by replacing the NO solve with a FE solve over $\Omega_{\mathrm{II}}$), which serves as the computational time benchmark throughout Section~\ref{sec:results}.
 
\begin{table}[t]
  \centering
  \caption{PI-DeepONet network dimensions for each loading regime.
           Layer dimensions are listed as $[\text{width} \times \text{depth},\,\ldots]$.
           Branch~1 is a multilayer perceptron (MLP) in all cases; Branch~2 uses a
           PointNet architecture in the dynamic regime only (Section~\ref{sec:arch}).
           The hyperelastic network uses the same branch and trunk dimensions as the
           static case but replaces strain boundary losses $\mathcal{L}_{\mathrm{bcs},\varepsilon}$
           with Cauchy stress boundary losses $\mathcal{L}_{\mathrm{bcs},\sigma}$.}
  \label{tab:arch}
  \begin{tabular}{@{}llll@{}}
    \toprule
    \textbf{Regime} & \textbf{Branch Net} & \textbf{Trunk Net} & \textbf{Activation} \\
    \midrule
    Static / quasi-static
      & $[200{\times}2,\ 100{\times}4,\ 800]$
      & $[2,\ 100{\times}4,\ 800]$
      & $\tanh$ \\
    Elasto-dynamic
      & PointNet $+\ [82{\times}4,\ 256,\ 800]$
      & $[2,\ 100{\times}4,\ 800]$
      & $\tanh$ \\
      & \quad and $[82{\times}4{\times}2,\ 100{\times}4,\ 800]$ & & \\
    \bottomrule
  \end{tabular}
\end{table}

\begin{algorithm}[H]
\SetAlgoLined
\caption{Schwarz alternating method at non-overlapping boundary in the coupling framework}
\label{algo_1}
\textbf{Initialization}: Set $\mathbf{u}_{NO}^{0} = \mathbf{0}$ in $\Omega_{II}$ and $\mathbf{u}_{FE}^{0} = \mathbf{0}$ in $\Omega_{I}$\;

\textbf{Main loop}:\\
\textbf{for} $n = 0 : n_{max} - 1$ \textbf{do}:\\
\hspace*{1em}Set $j = 0$,  $\epsilon$ a critical value, and relaxation parameter $\boldsymbol{\rho}_r = 0.5$ \\
\hspace*{1em}\textbf{while} $j \geq 0$ \textbf{do}:\\
\hspace*{2em} $j = j+1$\\
\hspace*{2em}\textbf{Model FE}:\\
\hspace*{3em} 1.  Receive the traction $\mathbf{T}_{|\Gamma_{I}^{in}}^{n, j-1}$  from Model NO and $\mathbf{u}_{|\Gamma_{I}^{out}}^{n,j-1}$ \\
\hspace*{4em}from the external sources.  \\
\hspace*{3em} 2. Solve $\mathbf{u}_{|\Omega_{I}}^{n,j}$ based on the boundary conditions and obtain $\boldsymbol{\sigma}_{|\Gamma_{I}^{in}}^{n,j}$.\\
\hspace*{3em} 3. Obtain the  $\mathbf{u}^{n, j}$  at $\Gamma_{I}^{in} $ and pass it to Model NO.\\

\hspace*{2em}\textbf{Model NO}:\\
\hspace*{3em} 1. Receive the interface information $\mathbf{u}_{|\Gamma_{II}}^{n, j}$ from Model FE.\\
\hspace*{3em} 2. Solve $\mathbf{u}_{|\Omega_{II}}^{n,j}$ based on the boundary conditions and obtain $\boldsymbol{\sigma}_{|\Gamma_{II}}^{n, j}$.\\
\hspace*{3em} 3. Calculate the relaxation formula: $\tilde{\boldsymbol{\sigma}}_{|\Gamma_{II}}^{n,j} = (1-\boldsymbol{\rho}_r)\boldsymbol{\sigma}_{|\Gamma_{II}}^{n,j} + \boldsymbol{\rho}_r\boldsymbol{\sigma}_{|\Gamma_{I}^{in}}^{n,j}$; 
\hspace*{4em} dot with the normal vector $\mathbf{n}_{|\Gamma^{in}_{I}}$ to obtain traction at interface: \\
\hspace*{5em}$\mathbf{T}_{|\Gamma_{I}^{in}}^{n, j} = \tilde{\boldsymbol{\sigma}}_{|\Gamma_{II}}^{n,j} \cdot \mathbf{n}_{|\Gamma^{in}_{I}}$; and pass it back to Model FE.\\
\hspace*{2em} \textbf{If} $\|\mathbf{u}^{n, j}_{|\Omega_{I}} - \mathbf{u}^{n, j-1}_{|\Omega_{I}}\|_{L^2} + \| \mathbf{u}^{n, j}_{|\Omega_{II}} - \mathbf{u}^{n, j-1}_{|\Omega_{II}}\|_{L^2}< \epsilon$, \textbf{end while}\\
\hspace*{1em} \textbf{End for}
\end{algorithm}

\section{Numerical Results and Discussion}
\label{sec:results}

\begin{figure}[!t]
    \centering
    \includegraphics[width=1\linewidth]{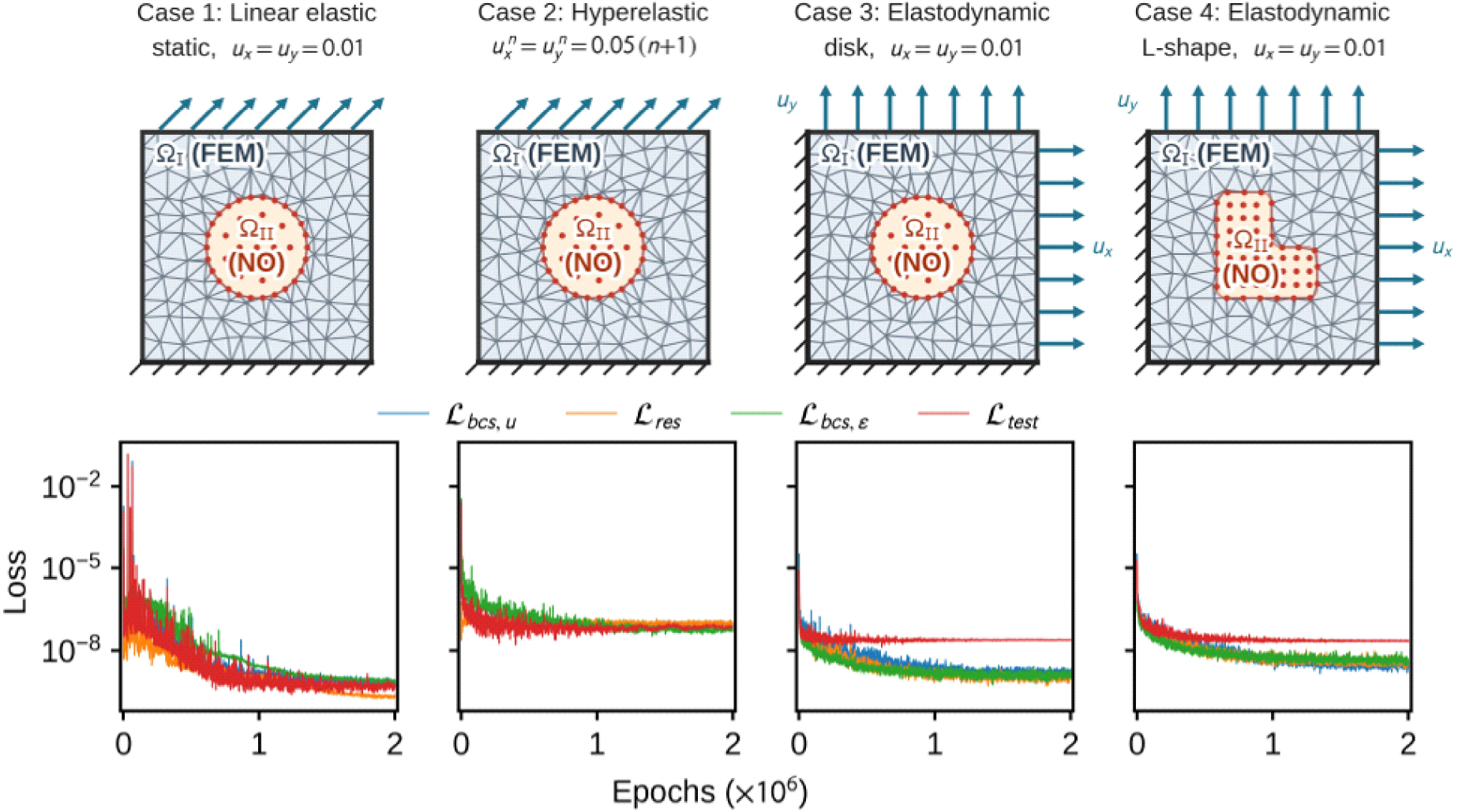}
    \caption{Schematics and training loss histories for the four validation cases. Cases $1$-$3$ share a square outer domain (side 2 units) with a circular inner subdomain $\Omega_{II}$ (radius $0.3$ units); Case 4 replaces the circle with a filleted L-shaped subdomain to demonstrate extension to irregular geometries. Loading increases in complexity from static linear elasticity (Case 1), to quasi-static hyperelasticity (Case 2), to elastodynamics with regular (Case 3) and irregular (Case 4) inner domains. All four training losses converge within $2\times10^6$ epochs, confirming the fidelity of the pretrained neural operators before coupling.}
    \label{fig:Schematics_loss}
\end{figure}
 
Four progressively complex cases validate the robustness and versatility of the FE-NO hybrid non-overlapping coupling framework presented in Section~\ref{sec:methodology}: (i) linear elasticity under static loading (Section~\ref{sec:static}), (ii) hyperelasticity under quasi-static loading (Section~\ref{sec:quasistatic}), and (iii) elastodynamics with regular circular inner domain Section~ \ref{sec:dynamic_disk}, and (iv) elastodynamics with irregular L-shaped inner domain (Section~\ref{sec:dynamic_lshape}). First three simulations, Cases~1-3, share the geometry of the outer and the inner subdomains, shown in Figure~\ref{fig:Schematics_loss}: a square domain of side~2 units with a circular inner subdomain $\Omega_{II}$ of radius~$0.3$ units, discretized into~$9{,}106$ triangular elements in $\Omega_{\mathrm{I}}$ (FE) and~$200$ uniformly distributed interface nodes on $\Gamma_{II} = \Gamma^{\mathrm{in}}_{\mathrm{I}}$. For Case~4, the inner domain is an L-shaped region with two arms - a long arm of length 1 unit and a short arm of length 0.5 units, whose six corners are rounded with fillets of radius 0.1 units, discretized into $10,850$  triangular elements in $ \Omega_I$ and $208$ uniformly distributed interface nodes on its shared interface. In the FE-FE benchmark, circular and L-shaped inner subdomains $\Omega_{II}$ are independently meshed with~$1{,}886$  and $4,958$ elements, respectively. Non-dimensionalised material properties are used throughout: density $\rho = 5$, time increment $\Delta t = 1$, Young's modulus $E = 1000$, and Poisson's ratio $\nu = 0.3$ (see Appendix~B). The two displacement DeepONets ($u_x$ and $u_y$) are trained with the Adam optimizer~\cite{kingma2014adam}; a learning rate of $\eta = 10^{-3}$ is used for both linear-elastic cases, and $\eta = 10^{-4}$ for the hyperelastic case owing to the stronger nonlinearity of its PDE residual.

\begin{figure}[H]
    \centering
    \includegraphics[width=1\linewidth]{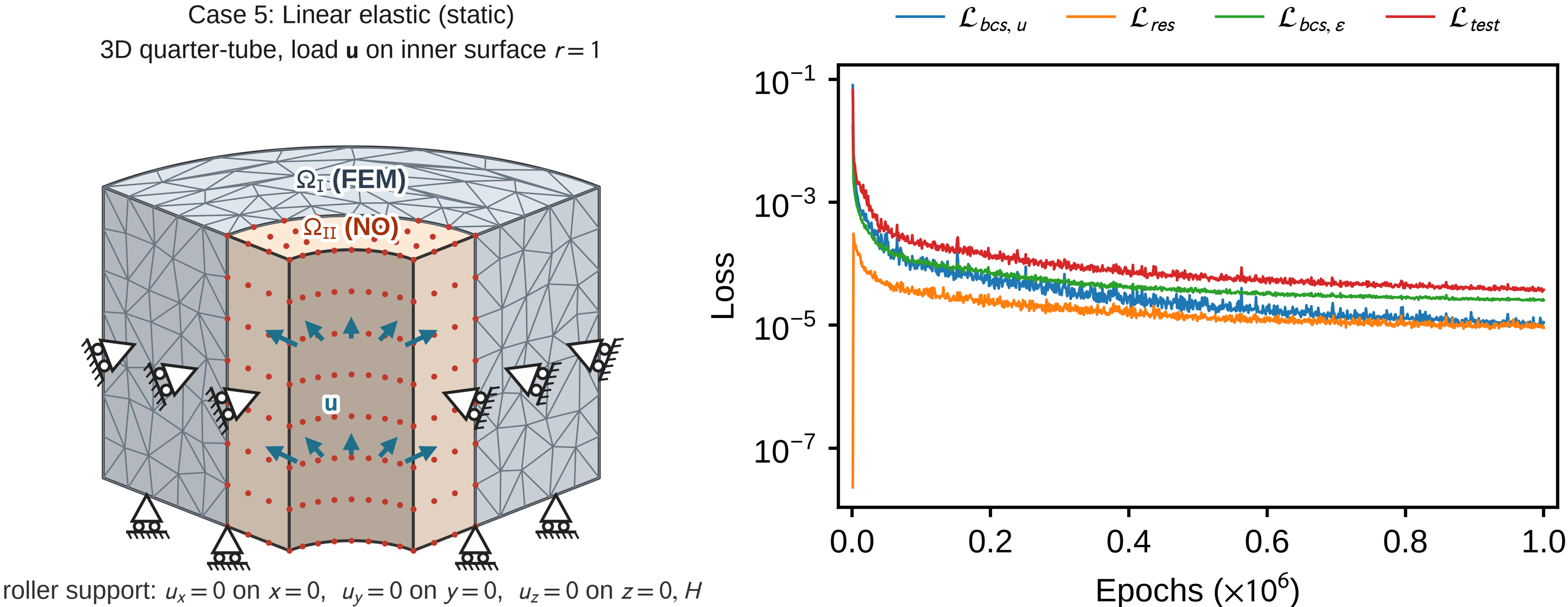}
    \caption{\WW{Schematics and training loss histories for the case 5. Left: The entire domain is a 4 unit high 3D quarter-tube with cylindrical surfaces at $r=1$  and $r = 4$. Inner and outer tube is decomposed at a cylindrical interface at r = 2.  A parabolic static load $\mathbf{u} = (u_x, u_y, u_z)= (0.01 \times x^2, 0.01 \times y^2, 0.05 \times z(z-4))$ is utilized with roller support at left, right, top and bottom boundaries.  Right: Training losses converge within $1\times10^6$ epochs, confirming the fidelity of the pretrained neural operators before coupling.}}
    \label{fig:Schematics_3D_tube}
\end{figure}

\WW{Besides the four 2D examples, a three-dimensional linear elastic quarter-tube case is presented in Section \ref{3D_tube} , with an inner quarter-tube ($\Omega_{II}$) and an outer tube ($\Omega_{I}$) as illustrated in Figure \ref{fig:Schematics_3D_tube}. $\Omega_{II}$  and $\Omega_{I}$  share an identical height $h = 4$  and a cylindrical interface $r=2$, with a inner and outer cylindrical surface at radius $r=1$ and $r=4$.  This quarter tube is subjected to a static loading $(u_x, u_y, u_z) = (0.01\times x^2, 0.01 \times y^2, 0.005\times z(z-4))$ at inner cylindrical surface with roller supports at all the flat boundaries in x, y, and z directions. The consequent high displacement concentration near the inner surface requires dense meshes, while the coarse meshes in the other regions are sufficient and computationally adequate. $\Omega_{I}$ and $\Omega_{II}$  are thus unevenly discretized into $28,845$ and $195,958$ elements, with degree of freedom (DOF) reaching 145, 995 and 859,857, respectively. To facilitate the information exchange at the cylindrical interface, uniformly distributed nodes ( $50$ in angular $ \times$  $80$  in height) are employed, providing the displacements ($15,741$ DOFs) as the input of Branch 1. To avoid GPU memory exhaustion, the MLP in Figure \ref{Fig:DeepONet_structure} is replaced by PointNet to efficiently encode the facial boundary conditions. }
 
A central advantage of the non-overlapping Neumann–Dirichlet scheme over the overlapping Dirichlet–Dirichlet framework of~\cite{wang2025time} is its markedly faster interface convergence. In the elastodynamic case, the number of inner Schwarz iterations per time step drops from 9–10 in the prior work to just 2–3 here, a reduction of more than threefold (see Table~\ref{tab:comparison}). This improvement stems from the complementary nature of the Neumann–Dirichlet transmission conditions, which enforce both traction continuity and displacement compatibility at the interface simultaneously, yielding a better-conditioned subdomain iteration than the symmetric Dirichlet–Dirichlet pairing. Analogous gains are observed across the static and quasi-static regimes, as detailed in Section~\ref{sec:dynamic}. \WW{The peak absolute errors in the FE-NO strain fields are $\mathcal{O}(10^{-3}-10^{-2})$, approximately one order of magnitude larger than the corresponding displacement errors. This increase is primarily attributable to two factors: (1) spatial differentiation of the predicted displacement fields, which can amplify existing approximation errors, particularly in regions with large displacement gradients; and (2) the difficulty of the fixed trunk bias in accurately representing sharp spatial features, as discussed in \cite{lanthaler2022error,kovachki2021universal}.}

\begin{table}[!t]
\centering
\caption{Quantitative comparison between the overlapping Dirichlet--Dirichlet
framework of~\cite{wang2025time} and the proposed non-overlapping
Neumann--Dirichlet framework for all validated benchmark cases. Note: Irregular subdomain geometry was not supported in the prior framework.}
\label{tab:comparison}
\footnotesize
\setlength{\tabcolsep}{6pt}
\renewcommand{\arraystretch}{1.15}
\begin{tabular}{@{}l l c c c c@{}}
\toprule
\multirow{2}{*}{\textbf{Case}} &
\multirow{2}{*}{\textbf{Method}} &
\multirow{2}{*}{$\boldsymbol{j_{\mathrm{cv}}}^*$} &
\multirow{2}{*}{$\boldsymbol{t_{\mathrm{wall}}^{**}}$\textbf{(s)}} &
\textbf{Max rel.} &
\textbf{Interp.} \\
& & & & \textbf{disp.\ err.\ (\%)} & \textbf{required?} \\
\midrule
\multirow{3}{*}{Static}
  & FE-FE (benchmark)              & 28      & 34.8  & $<$0.01 & No \\
  & FE-NO \cite{wang2025time}      & 11      & 22.2  & $<$0.5  & No \\
  & FE-NO, this work               & 10      & 12.8  & $<$0.2  & No \\
\midrule
\multirow{3}{*}{Quasi-static}
  & FE-FE (benchmark)              & 23      & 98    & $<$0.01 & No \\
  & FE-NO \cite{wang2025time}      & 50      & 160.8 & $<$2.5  & No \\
  & FE-NO, this work               & 16      & 56    & $<$0.5  & No \\
\midrule
\multirow{3}{*}{\shortstack[l]{Dynamic\\(regular)}}
  & FE-FE (disk, benchmark)        & 2--3    & 1.5   & $<$0.01 & No \\
  & FE-NO (sq.), \cite{wang2025time} & 9--10 & 188.8 & $<$2.5  & Yes (CNN) \\
  & FE-NO (disk, this work)        & 3--4    & 1.8   & $<$2.5  & No \\
\midrule
\multirow{2}{*}{\shortstack[l]{Dynamic\\(irregular)}}
  & FE-NO \cite{wang2025time}      & \multicolumn{4}{c}{N/A-CNN branch restricted to square domains} \\
  & FE-NO (L-shape, this work)     & 3       & 3.4   & $<$10   & No \\
\midrule
\multirow{2}{*}{\WW{Static (3D)}}
&\WW{ FE-FE (benchmark)}                 &\WW{25}     &\WW{1014.1} &  \WW{< 0.01}& \WW{No}\\
& \WW{FE-NO, this work}                      & \WW{24}     & \WW{175.8}   &  \WW{<  3.5} & \WW{No}\\
\midrule
\multicolumn{6}{@{}l@{}}{\footnotesize * $j_{\mathrm{cv}}$: inner Schwarz iterations to convergence.}\\
\multicolumn{6}{@{}l@{}}{\footnotesize ** $t_{\mathrm{wall}}$: wall-clock time per converged step.}\\
\bottomrule
\end{tabular}
\end{table}
\WW{To train the physics-informed neural operators, we generate boundary-condition/strain datasets for each benchmark problem using FE solver (no domain decomposition). For the static and quasi-static problems, 1,000 samples are generated for each case. The corresponding total data-generation times are 1,920~s for the 2D static case, 73,298~s for the quasi-static case, and 19,817~s for the 3D static quarter-tube case. For the elastodynamic problems, 100 independent loading samples are generated for each geometry, with each sample containing at least 10 time steps, resulting in more than 1,000 time-resolved training instances per geometry. The corresponding data-generation times are 2,108~s for the circular geometry and 3,146~s for the L-shaped geometry. These FE simulations constitute an offline preprocessing cost and are performed only once prior to training the neural operators.} 

\subsection{Case 1: Static Linear Elasticity}
\label{sec:static}

\begin{figure}[H]
\begin{center}
\includegraphics[width=1\textwidth]{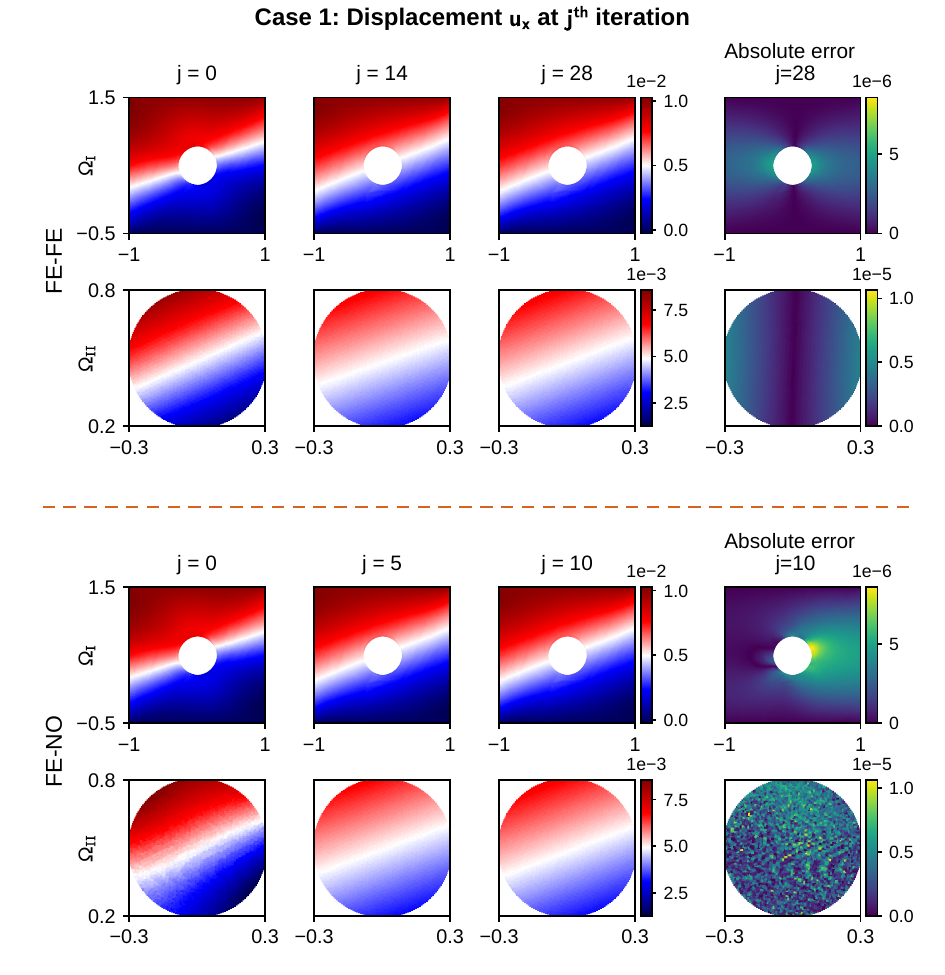}
\caption{Case~1: Schwarz iteration convergence of $u_x$ (static linear-elastic case).
           FE-FE (top two rows): $u_x$ in $\Omega_{\mathrm{I}}$ and $\Omega_{\mathrm{II}}$ at $j = 0, 14, 28$; FE-NO (bottom two rows):$u_x$ in $\Omega_{\mathrm{I}}$ and $\Omega_{\mathrm{II}}$ at $j = 0, 5, 10$; column~4 shows the absolute error $|u^{\mathrm{ref}}_x - u^{j_{\mathrm{cv}}}_x|$ at
           convergence. FE-NO reaches a visually converged state by $j = 10$; FE-FE requires $j = 28$. Maximum relative error at convergence: ${<}\,0.2\%$ for both schemes.}
\label{Fig:static_u}
\end{center}
\end{figure}

\begin{figure}[H]
\begin{center}
\includegraphics[width=1\textwidth]{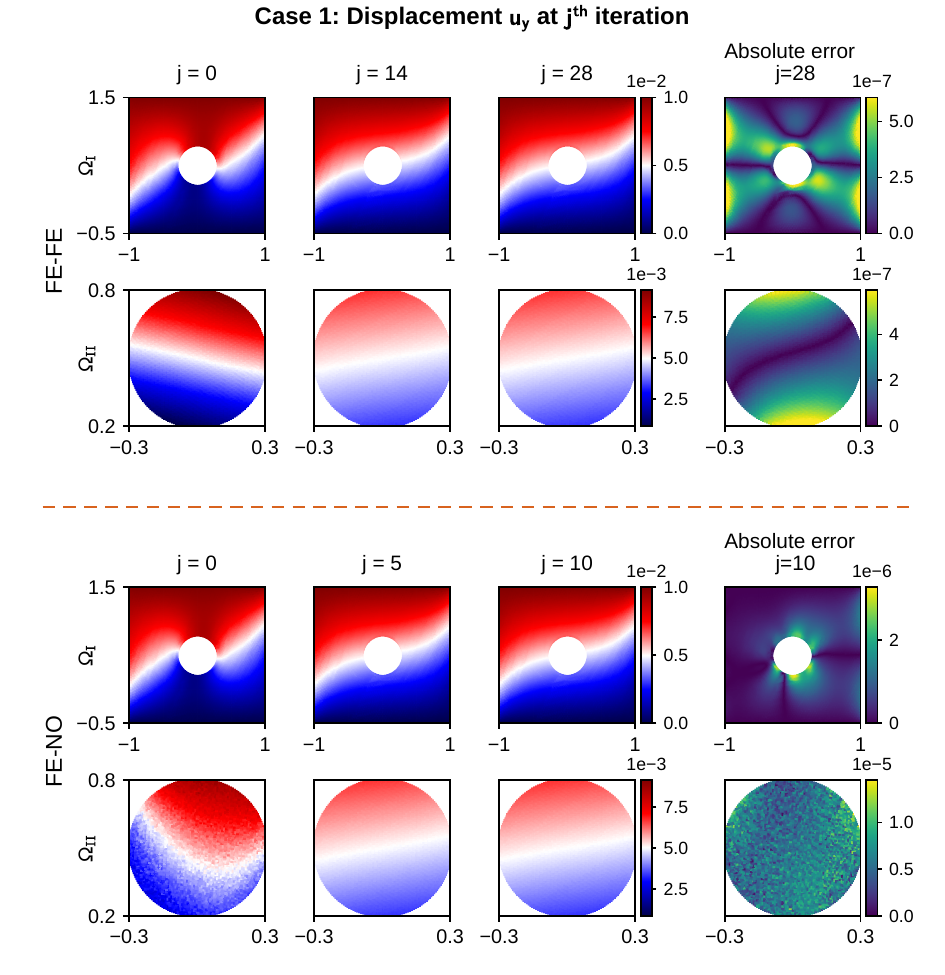}
\caption{Case~1: As Figure~\ref{Fig:static_u} but for $u_y$ (static case). Symmetric loading produces symmetric error fields in FE-FE and scattered errors in FE-NO, consistent with the mesh-free, data-driven nature of the pretrained operators. Maximum relative error at convergence: ${<}\,0.2\%$ for both schemes.}
\label{Fig:static_v}
\end{center}
\end{figure}

A linear elastic square is loaded statically: the bottom edge is fixed, and the top edge is subjected to uniform displacement $(u_x, u_y) = (0.01, 0.01)$ (Figure~\ref{fig:Schematics_loss} Case 1). The FE solution over the intact domain provides the ground truth fields $u^{\mathrm{ref}}_x$ and $u^{\mathrm{ref}}_y$ to calculate the absolute error in the last column of Figure \ref{Fig:static_u} - \ref{Fig:static_v}. Under static loading, Branch~2 (Point Net) is inactive; Branch~1 encodes only the Dirichlet boundary conditions $\mathbf{u}|_{\partial\Omega}$
(Section~\ref{sec:arch}). Substituting the displacement operators into Eqs.~\eqref{eq:static} and~\eqref{eq:strain_linear} yields the governing residuals over $\Omega_{II}$:
\begin{align}
  (\lambda + 2\mu)\frac{\partial^2 \mathcal{G}^{u_x}_{\bm\theta_1}}{\partial x^2}
  + \mu\frac{\partial^2 \mathcal{G}^{u_x}_{\bm\theta_1}}{\partial y^2}
  + (\lambda + \mu)\frac{\partial^2 \mathcal{G}^{u_y}_{\bm\theta_2}}{\partial x\,\partial y}
  &= 0, \quad (x,y)\in\Omega_{II}, \label{eq:static_res_x} \\
  (\lambda + 2\mu)\frac{\partial^2 \mathcal{G}^{u_y}_{\bm\theta_2}}{\partial y^2}
  + \mu\frac{\partial^2 \mathcal{G}^{u_y}_{\bm\theta_2}}{\partial x^2}
  + (\lambda + \mu)\frac{\partial^2 \mathcal{G}^{u_x}_{\bm\theta_1}}{\partial x\,\partial y}
  &= 0, \quad (x,y)\in\Omega_{II}, \label{eq:static_res_y}
\end{align}
with Dirichlet boundary conditions on $\Gamma_{II}$:
\begin{align}
  \mathcal{G}^{u_x}_{\bm\theta_1} = u_x(x,y), \quad
  \mathcal{G}^{u_y}_{\bm\theta_2} = u_y(x,y), \quad (x,y)\in\Gamma_{II},
  \label{eq:static_bc_u}
\end{align}
and strain boundary conditions (Eqs.~\eqref{eq:eps_xx}-\eqref{eq:eps_yy}):
\begin{equation}
  \mathcal{G}^{\varepsilon_{xx}}_{\bm\theta_1,\bm\theta_2} = \varepsilon_{xx},\quad
  \mathcal{G}^{\varepsilon_{xy}}_{\bm\theta_1,\bm\theta_2} = \varepsilon_{xy},\quad
  \mathcal{G}^{\varepsilon_{yy}}_{\bm\theta_1,\bm\theta_2} = \varepsilon_{yy},
  \quad (x,y)\in\Gamma_{II}.
  \label{eq:static_bc_eps}
\end{equation}
Boundary data for Eqs.~\eqref{eq:static_bc_u}-\eqref{eq:static_bc_eps} are generated via GRF sampling (Appendix~A). After $2\times10^6$ training epochs, the test loss evaluated on five unseen samples converges to
$5\times10^{-10}$ (Figure~\ref{fig:Schematics_loss} Case 1 bottom row), confirming the good training of the neural operators.

\begin{figure}[H]
\begin{center}
\includegraphics[width=1\textwidth]{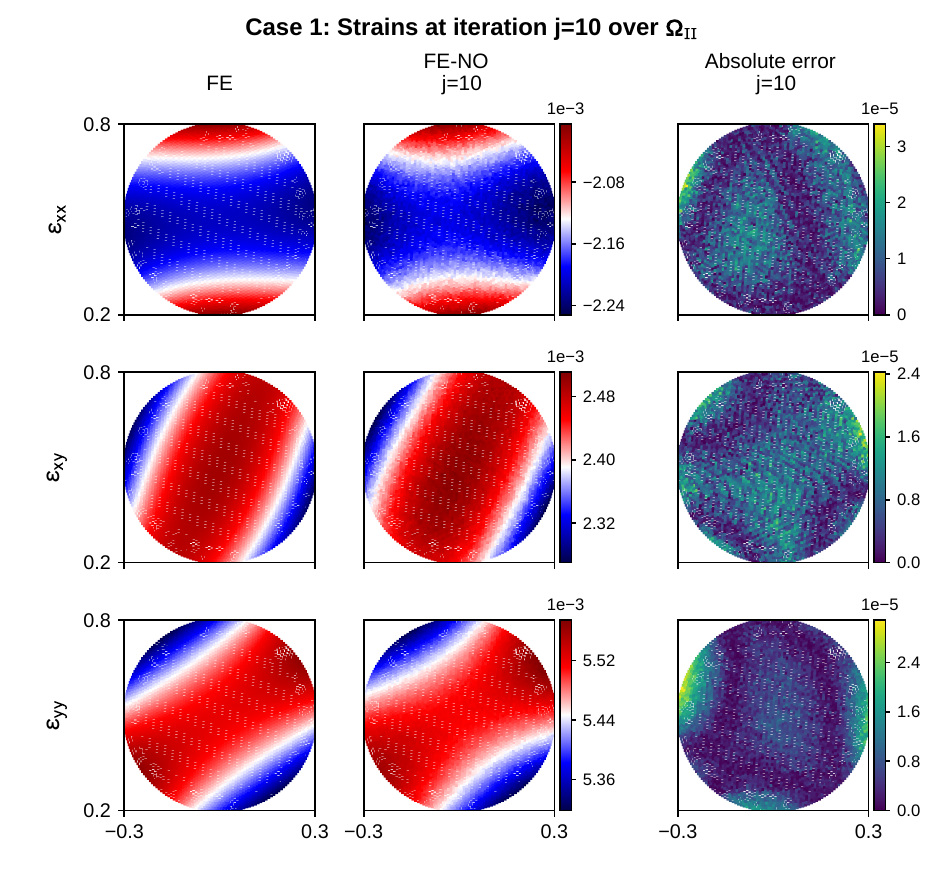}
\caption{Case~1: Converged strain fields from FE-NO at $j = 10$ (static case):
           FE reference (column~1), FE-NO prediction (column~2), and absolute
           error (column~3) for $\varepsilon_{xx}$, $\varepsilon_{xy}$, and
           $\varepsilon_{yy}$.
           Strain operators are derived analytically from the displacement
           operators via Eqs.~\eqref{eq:eps_xx}-\eqref{eq:eps_yy}; the maximum relative strain error is $1.5\%$.}
\label{Fig:strains_LE}
\end{center}
\end{figure}

\begin{figure}[H]
\begin{center}
\includegraphics[width=1\textwidth]{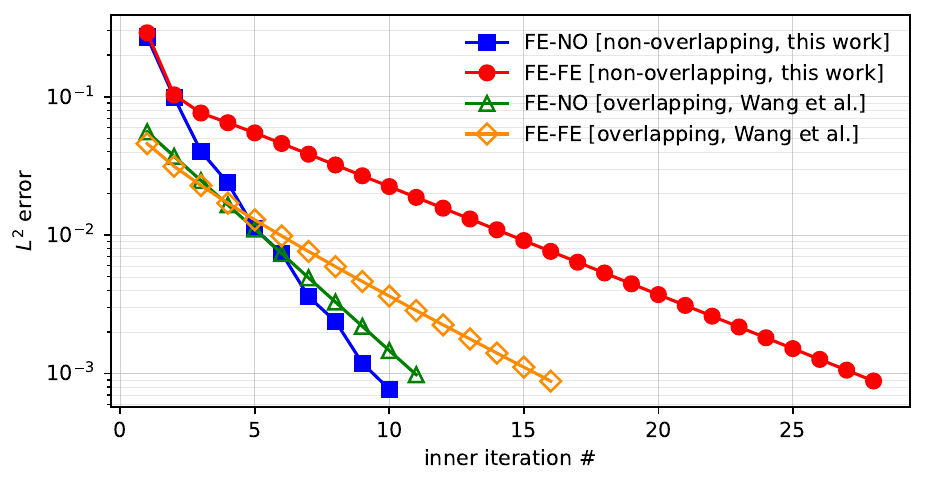}
\caption{Case~1: $L^2$ error norm (Eq.~\eqref{eq:convergence}) versus inner Schwarz
           iteration in our current and prior work \cite{wang2025time}. 
           In this non-overlapping work, FE-NO converges in 10 iterations; FE-FE requires 28.
           The divergence after iteration~2 reflects continuous stress output
           from the mesh-free NO versus mesh-induced stress discontinuities in
           FE-FE. Wall-clock times: FE-NO $12.8\,\mathrm{s}$,FE-FE $34.8\,\mathrm{s}$ ($2.72{\times}$ faster. In prior overlapping work, FE-NO and FE-FE require 11 and 16 iterations, respectively. Wall-clock times: FE-NO $22.2\,\mathrm{s}$, FE-FE $55.2\,\mathrm{s}$.}
\label{Fig:static_error}
\end{center}
\end{figure}

\subsubsection{Coupling results}
\label{sec:static_results}
 
Figures~\ref{Fig:static_u} and~\ref{Fig:static_v} show the evolution of $u^j_x$ and $u^j_y$ across inner Schwarz iterations~$j$ in both coupling frameworks (FE-FE and FE-NO); the superscript here indexes inner iterations, not time steps. At $j = 0$ both frameworks initialize with a uniform-zero field in $\Omega_{II}$. By $j = 5$, FE-NO reaches a visually converged state; FE-FE requires~$j = 14$ before a comparable distribution is achieved, and does not fully converge until~$j = 28$. The absolute displacement error at convergence relative to the FE ground truth is below~$0.2\%$ in all domains and both frameworks, confirming the accuracy of the Neumann-Dirichlet approach. Although the FE-FE error map is symmetric (consistent with the symmetric geometry and loading), the FE-NO error in $\Omega_{II}$ is scattered rather than symmetric, reflecting the mesh-free, data-driven nature of the pretrained operators.
 
Figure~\ref{Fig:strains_LE} presents the converged strain fields $(\varepsilon^{10}_{xx}, \varepsilon^{10}_{xy}, \varepsilon^{10}_{yy})$ from FE-NO alongside their absolute errors. The analytically derived strain operators (Eqs.~\eqref{eq:eps_xx}-\eqref{eq:eps_yy}) produce speckled error patterns with a diffuse interface, yet the maximum relative strain error does not exceed~$1.5\%$, validating the mechanically consistent derivation. The convergence behaviour is quantified in Figure~\ref{Fig:static_error}. The $L^2$ error norms of FE-FE and FE-NO overlap for the first two iterations, after which they diverge sharply: FE-NO decreases monotonically to the threshold $\varepsilon = 10^{-3}$ in~10 iterations, whereas FE-FE stalls with a reduced descent rate and requires~28 iterations. This difference originates from the Dirichlet data passed to $\Omega_{II}$: when the FE solver in $\Omega_{II}$ receives a Dirichlet condition at a mesh-constrained boundary, stress discontinuities develop near the interface that slow convergence. The mesh-free DeepONet, by contrast, always returns a continuous stress field, sustaining the rapid convergence rate.
 
As a direct consequence, the FE-NO coupling completes in~$12.8\,\mathrm{s}$ versus~$34.8\,\mathrm{s}$ for FE-FE, enabling a~$172\%$ reduction in wall-clock time. Compared to the overlapping Dirichlet-Dirichlet scheme of~\cite{wang2025time}, which required~11 inner iterations per solve ($22.2\,\mathrm{s}$), the non-overlapping Neumann-Dirichlet coupling achieves the same accuracy while being $1.73 \times$ faster, establishing its efficiency advantage from the outset.

\subsection{Case 2: Quasi-Static Hyperelasticity}
\label{sec:quasistatic}

A Neo-Hookean square undergoes progressively increasing displacement loading: the bottom edge is fixed, and the top edge displaces as $(u^n_x, u^n_y) = (0.05(n+1),\, 0.05(n+1))$ over five time steps $n = 0,\ldots,4$ (Figure~\ref{fig:Schematics_loss} Case 2). The FE solution over the intact domain at $n = 4$ serves as the ground truth. For the hyperelastic PI-DeepONet, Branch~1 encodes Dirichlet boundary displacements. Substituting the displacement operators into the large-deformation equilibrium (Eqs.~\eqref{eq:EGL}-\eqref{eq:qs_equil})
yields the nonlinear residuals:
\begin{align}
  &\frac{\partial}{\partial x}\!\left[
    \mu\!\left(1 + \frac{\partial \mathcal{G}^{u_x}_{\bm\theta_1}}{\partial x}\right)
    + \left(\lambda\ln J - \mu\right)
      \frac{1}{J}\!\left(1 + \frac{\partial \mathcal{G}^{u_y}_{\bm\theta_2}}{\partial y}\right)
  \right] \nonumber \\
  &+\frac{\partial}{\partial y}\!\left[
    \mu\frac{\partial \mathcal{G}^{u_y}_{\bm\theta_2}}{\partial x}
    + \left(\lambda\ln J - \mu\right)
      \frac{1}{J}\!\left(-\frac{\partial \mathcal{G}^{u_x}_{\bm\theta_1}}{\partial y}\right)
  \right] = 0, \quad (x,y)\in\Omega_{II},
  \label{eq:qs_res_x} \\[4pt]
  &\frac{\partial}{\partial y}\!\left[
    \mu\!\left(1 + \frac{\partial \mathcal{G}^{u_y}_{\bm\theta_2}}{\partial y}\right)
    + \left(\lambda\ln J - \mu\right)
      \frac{1}{J}\!\left(1 + \frac{\partial \mathcal{G}^{u_x}_{\bm\theta_1}}{\partial x}\right)
  \right] \nonumber \\
  &+\frac{\partial}{\partial x}\!\left[
    \mu\frac{\partial \mathcal{G}^{u_x}_{\bm\theta_1}}{\partial y}
    + \left(\lambda\ln J - \mu\right)
      \frac{1}{J}\!\left(-\frac{\partial \mathcal{G}^{u_y}_{\bm\theta_2}}{\partial x}\right)
  \right] = 0, \quad (x,y)\in\Omega_{II},
  \label{eq:qs_res_y}
\end{align}
where the 2-D Jacobian determinant is
\begin{equation}
  J = \left(1 + \frac{\partial \mathcal{G}^{u_x}_{\bm\theta_1}}{\partial x}\right)
      \left(1 + \frac{\partial \mathcal{G}^{u_y}_{\bm\theta_2}}{\partial y}\right)
    - \frac{\partial \mathcal{G}^{u_x}_{\bm\theta_1}}{\partial y}
      \frac{\partial \mathcal{G}^{u_y}_{\bm\theta_2}}{\partial x}.
  \label{eq:J_2d}
\end{equation}
Because the Cauchy stress in large deformation cannot be expressed as a linear function of strain (cf.\ Eqs.~\eqref{eq:EGL}-\eqref{eq:Cauchy_large}), the strain boundary operators of the static case are replaced by Cauchy stress operators $\mathcal{G}^{\sigma_{xx}}_{\bm\theta_1,\bm\theta_2}$, $\mathcal{G}^{\sigma_{xy}}_{\bm\theta_1,\bm\theta_2}$, and $\mathcal{G}^{\sigma_{yy}}_{\bm\theta_1,\bm\theta_2}$, derived from $\{\mathcal{G}^{u_x}_{\bm\theta_1},\mathcal{G}^{u_y}_{\bm\theta_2}\}$ via Eqs.~\eqref{eq:PK2}-\eqref{eq:Cauchy_large}:
\begin{equation}
  \mathcal{G}^{\sigma_{ij}}_{\bm\theta_1,\bm\theta_2} = \sigma_{ij}(x,y),
  \quad ij\in\{xx,xy,yy\},\quad (x,y)\in\Gamma_{II}.
  \label{eq:qs_bc_sig}
\end{equation}
All boundary data (Eqs.~\eqref{eq:static_bc_u} and~\eqref{eq:qs_bc_sig}) are GRF-sampled (Appendix~A). After $2\times10^6$ training iterations, the test loss converges to $\mathcal{O}(10^{-8})$ (Figure~\ref{fig:Schematics_loss} Case~2), confirming the fidelity of the pretrained operators for this strongly nonlinear problem.

\begin{figure}[H]
\begin{center}
\includegraphics[width=1\textwidth]{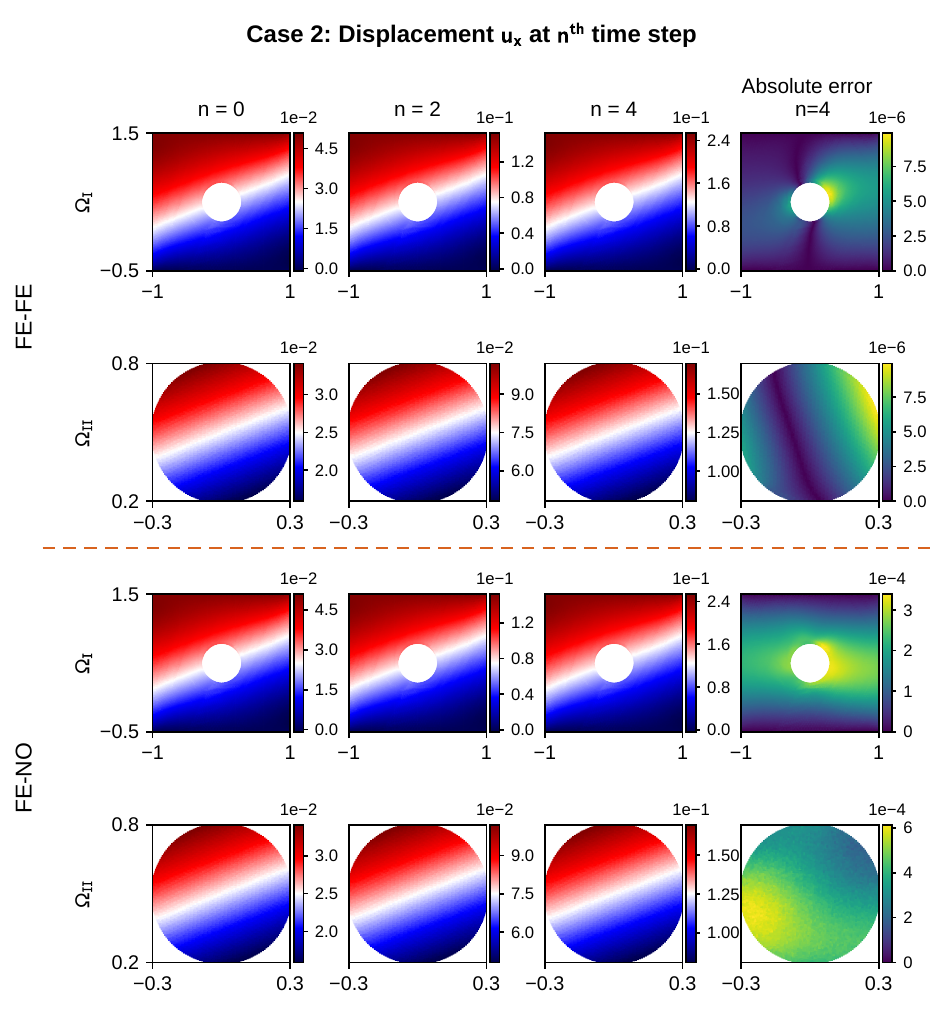}
\caption{Case~2: Quasi-static hyperelastic coupling  -  $u_x$ at time steps $n = 0, 2, 4$ for FE-FE (top two rows) and FE-NO (bottom two rows); within each coupling method, two rows correspond to subdomains $\Omega_{\mathrm{I}}$  and $\Omega_{\mathrm{II}}$; column~4 shows the absolute error at $n = 4$. Peak field magnitude grows from $5{\times}10^{-2}$ to $0.25$; FE-NO absolute error $\mathcal{O}(10^{-4})$, FE-FE error $\mathcal{O}(10^{-6})$.}
\label{Fig:quasi_static_u}
\end{center}
\end{figure}

\begin{figure}[H]
\begin{center}
\includegraphics[width=1\textwidth]{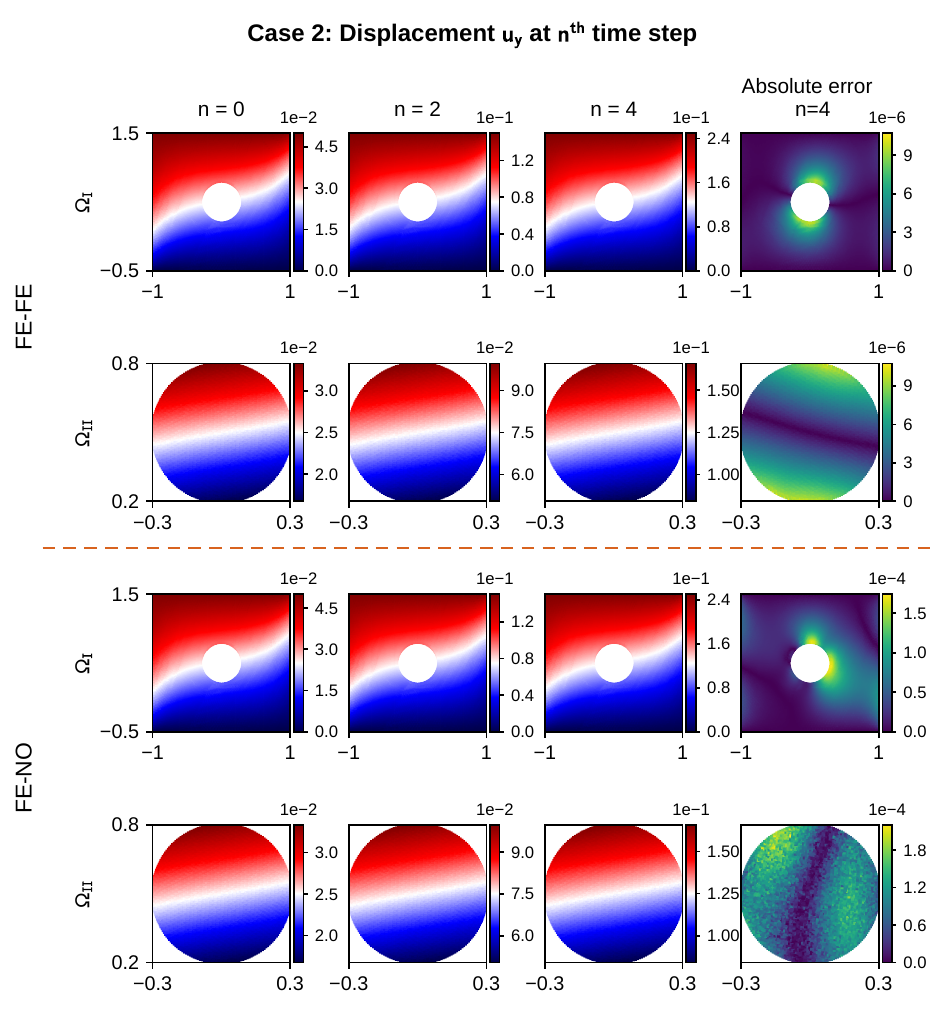}
\caption{Case~2: As Figure~\ref{Fig:quasi_static_u} but for $u_y$ (quasi-static case). An asymmetric error distribution emerges in FE-FE despite the
           symmetric geometry and loading, arising from Newton-Raphson
           nonlinearity and the morphological evolution of the Neumann interface
           normal under finite deformation
           (Eqs.~\eqref{eq:nanson}-\eqref{eq:normal_large}).
           FE-NO errors are larger in magnitude but distributed more uniformly.}
\label{Fig:quasi_static_v}
\end{center}
\end{figure}

\begin{figure}[H]
\begin{center}
\includegraphics[width=1\textwidth]{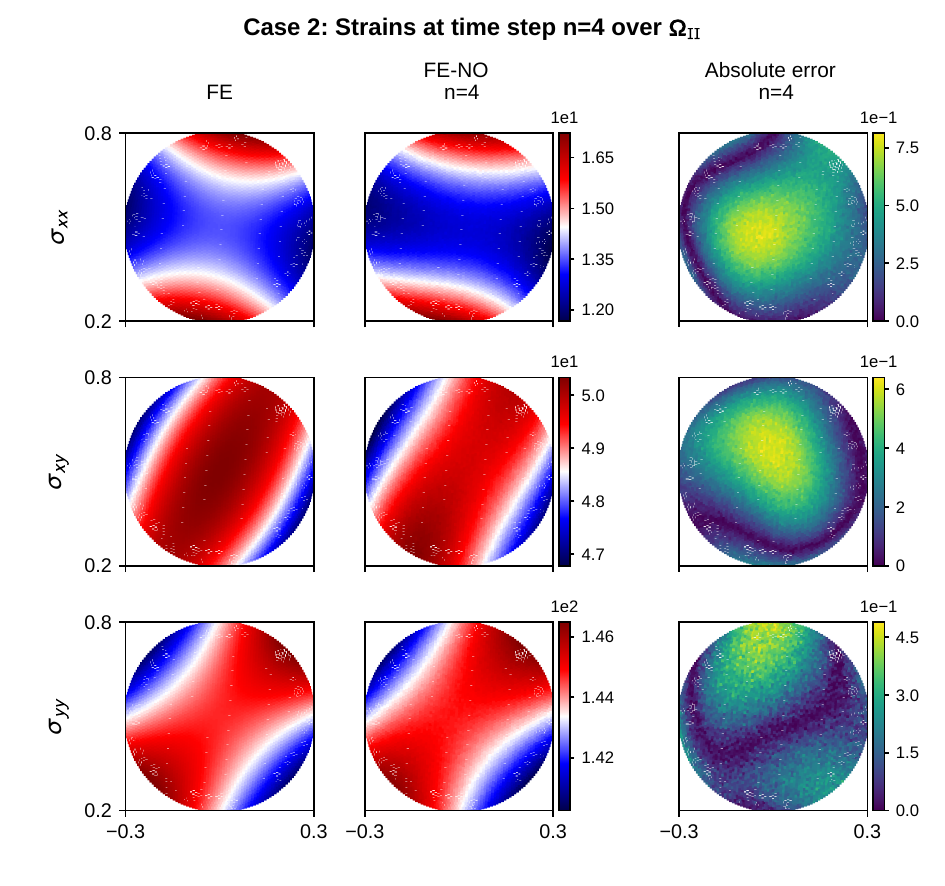}
\caption{Case~2: Converged Cauchy stress fields in $\Omega_{\mathrm{II}}$ at $n = 4$
           (quasi-static hyperelastic case):
           FE reference (column~1), FE-NO prediction (column~2), and absolute
           error (column~3) for $\sigma_{xx}$, $\sigma_{xy}$, and $\sigma_{yy}$.
           Maximum absolute errors are $\mathcal{O}(0.8)$, negligible relative
           to the stress magnitude, confirming the accuracy of the Schwarz
           alternating method for strongly nonlinear problems.}
\label{Fig:stress_HE}
\end{center}
\end{figure}

\begin{figure}[H]
\begin{center}
\includegraphics[width=1\textwidth]{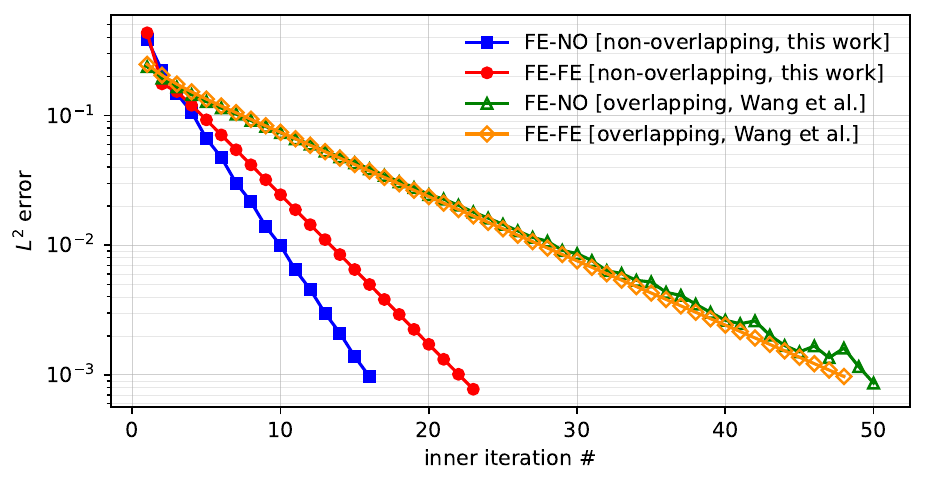}
\caption{Case~2: $L^2$ error norm versus inner Schwarz iteration at $n = 4$
           (quasi-static case) in our current and prior work \cite{wang2025time}.
           In this non-overlapping work, FE-NO converges in 16 iterations; FE-FE requires 23 - a pattern
           consistent across all steps: $j = \{18,20,22,24,16\}$ for FE-NO
           and $\{24,25,27,29,23\}$ for FE-FE at $n = \{0,1,2,3,4\}$.
           Wall-clock times at $n = 4$: FE-NO $56\,\mathrm{s}$,
           FE-FE $98\,\mathrm{s}$ ($1.75{\times}$ faster). In prior overlapping work, FE-NO and FE-FE require $50$ and $48$ iterations to converge, respectively. Wall-clock time: FE-NO $160.8\,\mathrm{s}$  and FE-FE $184.8\,\mathrm{s}$.}
\label{Fig:quasi_static_hyper_error_profile}
\end{center}
\end{figure}

\subsubsection{Coupling results}
\label{sec:qs_results}
 
The Schwarz alternating scheme of Algorithm~\ref{algo_1} is applied identically to the quasi-static case, with the outer time loop advancing the applied displacement at each step $n$. The convergence criterion and relaxation parameter ($\varepsilon = 10^{-3}$, $\rho_r = 0.5$) are inherited from the previous example, and only the converged fields at each $n$ are reported.
 
Figures~\ref{Fig:quasi_static_u} and~\ref{Fig:quasi_static_v} show the evolution of $u^n_x$ and $u^n_y$ at $n = 0, 2, 4$ for both coupling frameworks (FE-FE and FE-NO). In $\Omega_{I}$, both schemes produce fields that grow uniformly from a maximum of~$5\times10^{-2}$ at $n = 0$ to~$0.25$ at $n = 4$. The absolute error $|u^4_{x,\mathrm{FE}} - u^4_{x,\mathrm{FE\text{-}NO}}|$ reaches~$\mathcal{O}(10^{-4})$, approximately~100 times larger than the FE-FE error ($\mathcal{O}(10^{-6})$), yet remains negligible relative to the displacement magnitude. An asymmetric error distribution emerges in $u^4_{x,\mathrm{FE\text{-}FE}}$ and $u^4_{y,\mathrm{FE\text{-}FE}}$ despite the symmetric geometry and loading - a consequence of the strongly nonlinear Newton-Raphson solver~\cite{petsc-web-page} and the morphological evolution of the Neumann interface under finite deformation (Eqs.~\eqref{eq:nanson}-\eqref{eq:normal_large}). The FE-NO error, while larger in magnitude, is distributed more uniformly.
 
The converged Cauchy stress fields at $n = 4$ are shown in Figure~\ref{Fig:stress_HE}. All three components agree closely with the FE reference, with maximum absolute errors of~$\mathcal{O}(0.8)$, negligible relative to the stress magnitude. These results confirm that the Schwarz alternating method extends reliably to strongly nonlinear hyperelastic problems.
 
\paragraph{Convergence efficiency}
Figure~\ref{Fig:quasi_static_hyper_error_profile} presents the $L^2$ error profiles at $n = 4$. FE-NO converges in~16 inner iterations; FE-FE requires~23, a pattern that persists across all time steps ($j = \{18,20,22,24,16\}$ for FE-NO and $\{24,25,27,29,23\}$ for FE-FE at $n = \{0,1,2,3,4\}$). The monotonic decrease in $j$ with advancing $n$ reflects the auto-regressive algorithm's ability to leverage the kinetic history of the previous step, progressively easing inter-domain communication. This capacity becomes critical in the next example, where the auto-regressive loop runs for hundreds of steps.
 
\paragraph{Computational speed} At $n = 4$, FE-NO completes in~$56\,\mathrm{s}$ versus~$98\,\mathrm{s}$ for FE-FE, a $1.75\times$ speed up is achieved. Two factors drive this gain. First, the lower iteration count reduces the number of FE solves in $\Omega_I$. Second, the Newton-Raphson linearisation required by the FE solver in $\Omega_{II}$ for each inner iteration $j$ is avoided entirely: the pretrained NO evaluates $\Omega_{II}$ in a single forward pass without recalculation. This latter advantage also enhances robustness, since conventional nonlinear FE solvers can fail under large load increments or severe element distortion, whereas the NO is fundamentally mesh-free and designed for nonlinear mappings.
 

Compared to the overlapping Dirichlet-Dirichlet scheme of~\cite{wang2025time}, the present non-overlapping approach, which exchanges interface information through a Neumann–Dirichlet condition, requires less than half as many inner iterations: the count drops from $48$ to $16$ for FE-NO and from $50$ to $23$  for FE-FE. Consequently, it achieves the wall-clock speed-ups of  $2.87\times$ and $1.88\times$ for FE-NO and FE-FE, respectively.

\subsection{Case 3 and Case 4: elastodynamics}
\label{sec:dynamic}

Two linear elastic squares are loaded identically in the dynamic regime: the left and bottom edges are fixed, a constant uniform displacement $u_x = 0.01$ is applied to the right edge, and $u_y = 0.01$ to the top edge (Figure~\ref{fig:Schematics_loss} Case~3 and Case~4). The only difference between the two sub-cases is the geometry of $\Omega_{II}$: a circular disk (Section~\ref{sec:dynamic_disk}) or a filleted L-shaped domain (Section~\ref{sec:dynamic_lshape}).
 
Time-dependence requires both branch networks to be active (Section~\ref{sec:arch}): Branch~1 (MLP) encodes the current-step boundary conditions $(u^n|_{\Gamma_{II}},\,\varepsilon^n|_{\Gamma_{II}})$, and Branch~2 extracts the prior-step kinetic state $(u^{n-1}|_{\Omega_{II}},\,\dot{u}^{n-1}|_{\Omega_{II}})$. The governing residuals at time step~$n$ are obtained by substituting the displacement operators into the effective dynamic system Eq.~\eqref{eq:effective_system}:
\begin{align}
  (\lambda+2\mu)\frac{\partial^2 \mathcal{G}^{u_x}_{\bm\theta_1}}{\partial x^2}
  + \mu\frac{\partial^2 \mathcal{G}^{u_x}_{\bm\theta_1}}{\partial y^2}
  + (\lambda+\mu)\frac{\partial^2 \mathcal{G}^{u_y}_{\bm\theta_2}}{\partial x\partial y}
  &= \frac{-2}{(\beta\Delta t)^2}
     \!\left(u^{n-1}_x + \dot{u}^{n-1}_x\Delta t - \mathcal{G}^{u_x}_{\bm\theta_1}\right),
  \label{eq:dyn_res_x} \\
  (\lambda+2\mu)\frac{\partial^2 \mathcal{G}^{u_y}_{\bm\theta_2}}{\partial y^2}
  + \mu\frac{\partial^2 \mathcal{G}^{u_y}_{\bm\theta_2}}{\partial x^2}
  + (\lambda+\mu)\frac{\partial^2 \mathcal{G}^{u_x}_{\bm\theta_1}}{\partial x\partial y}
  &= \frac{-2}{(\beta\Delta t)^2}
     \!\left(u^{n-1}_y + \dot{u}^{n-1}_y\Delta t - \mathcal{G}^{u_y}_{\bm\theta_2}\right),
  \label{eq:dyn_res_y}
\end{align}
for $(x,y)\in\Omega_{II}$, with boundary conditions
\begin{equation}
  \mathcal{G}^{u_x}_{\bm\theta_1} = u^n_x,\quad
  \mathcal{G}^{u_y}_{\bm\theta_2} = u^n_y,\quad
  \mathcal{G}^{\varepsilon_{ij}}_{\bm\theta_1,\bm\theta_2} = \varepsilon^n_{ij},
  \quad ij\in\{xx,xy,yy\},\quad (x,y)\in\Gamma_{II}.
  \label{eq:dyn_bc}
\end{equation}
 
\paragraph{Training data} Boundary conditions $\mathbf{u}^n|_{\Gamma_{II}}$ and $\bm{\varepsilon}^n|_{\Gamma_{II}}$, and domain kinetic fields $\mathbf{u}^{n-1}|_{\Omega_{II}}$ and $\dot{\mathbf{u}}^{n-1}|_{\Omega_{II}}$ are generated by FE with GRF sampling (Appendix~A). A key advantage of Branch~2 is that these kinetic fields can be ingested directly from the unstructured FE mesh, without the interpolation step required by the CNN used in~\cite{wang2025time}. Furthermore, whereas the CNN architecture of~\cite{wang2025time} restricted the inner domain to a square (due to the requirement for regular grids), the PointNet operates on arbitrary point clouds, enabling the filleted L-shaped geometry in Section~\ref{sec:dynamic_lshape} and, in principle, three-dimensional domains~\cite{qi2017pointnet}.
 
Each pretrained NO covers a fixed window of time steps: 11 steps for the
disk geometry and 80 steps for the L-shape, with successive NOs
employed auto-regressively beyond those windows. After $2\times10^6$
training iterations, the test loss converges to
$2\times10^{-8}$ for both geometries (Figure~\ref{fig:Schematics_loss} Case 3 and Case 4),
demonstrating reliable generalization to unseen dynamic conditions.
 
\subsubsection{Case 3: Regular inner domain (circular disk)}
\label{sec:dynamic_disk}
 
Ground truth fields $u^{149}_{x,\mathrm{FE}}$ and
$u^{149}_{y,\mathrm{FE}}$ from the intact FEniCSx model serve as
reference. All kinetic quantities in $\Omega_{II}$ at $n = 98$ are
initialized from FE-FE results to provide consistent starting conditions
for both coupling schemes.

 \begin{figure}[H]
\begin{center}
\includegraphics[width=1\textwidth]{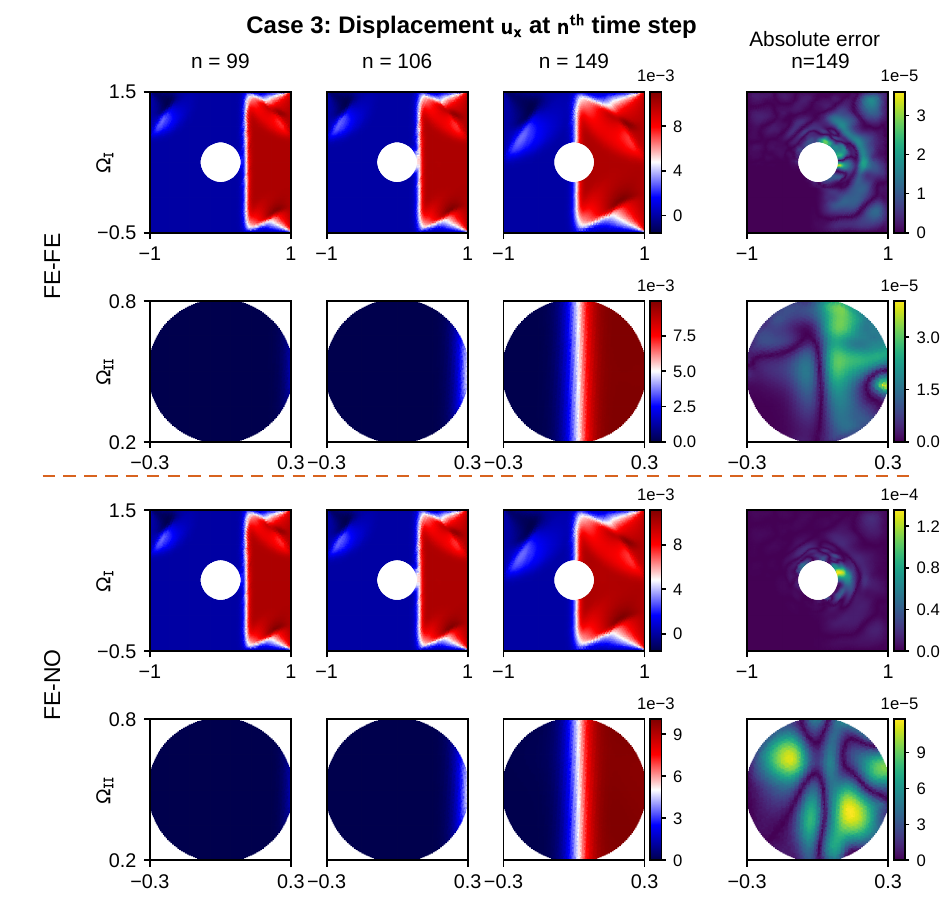}
\caption{Case~3 (Disk Example): Dynamic coupling - $u_x$ wave propagation at $n = 99, 106, 149$ for FE-FE (top two rows) and FE-NO (bottom two rows); within each coupling method, two rows corresponds to subdomains $\Omega_{\mathrm{I}}$ and $\Omega_{\mathrm{II}}$; the horizontal plane wave enters $\Omega_{\mathrm{II}}$ by $n = 106$ and fills it till $n = 149$ with no visible interface mismatch.
           Final absolute errors: $\mathcal{O}(10^{-5})$ for FE-FE and
           $\mathcal{O}(10^{-4})$ for FE-NO, both ${<}\,0.1\%$ relative to
           field magnitudes of $\mathcal{O}(10^{-1})$.}
\label{Fig:elasto_dynamic_u}
\end{center}
\end{figure}

\begin{figure}[H]
\begin{center}
\includegraphics[width=1\textwidth]{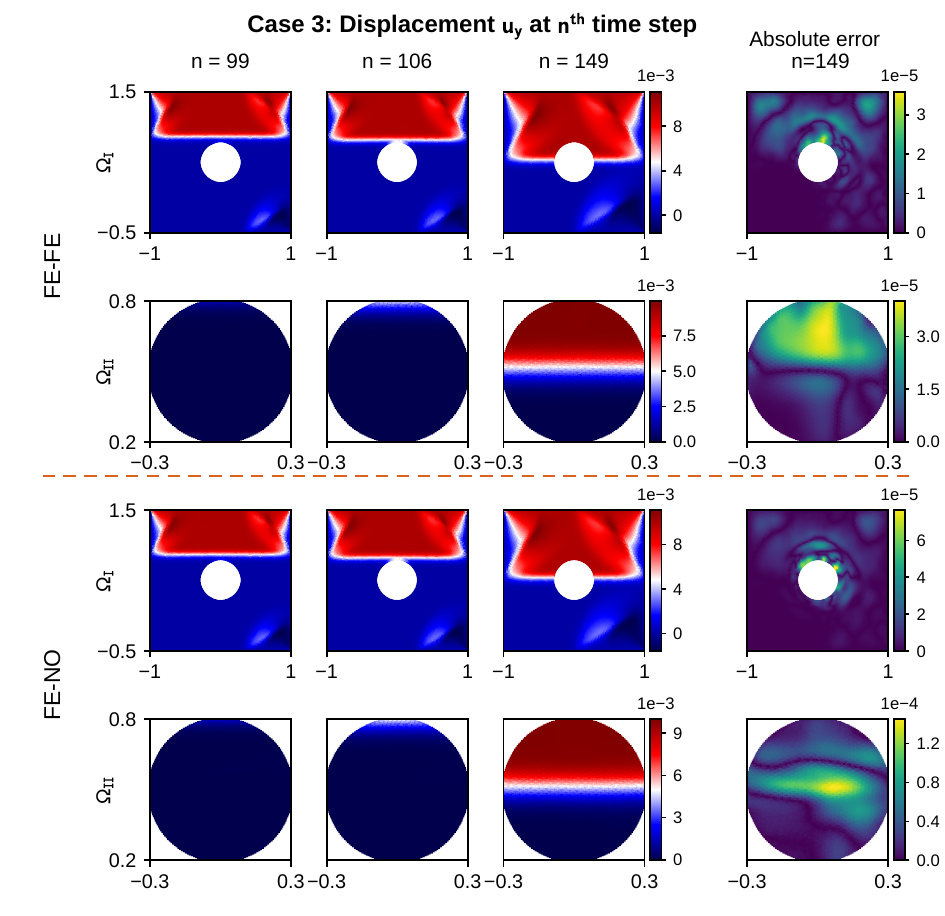}
\caption{Case~3 (Disk Example): As Figure~\ref{Fig:elasto_dynamic_u} but for $u_y$ (disk case). The vertical plane wave propagates downward; peak errors concentrate
           near the red-to-blue transition interface, consistent with the known
           difficulty of approximating sharp spatial gradients in operator-learning
           frameworks.}
\label{Fig:elasto_dynamic_v}
\end{center}
\end{figure}
 
\paragraph{Displacement fields and errors}
Figures~\ref{Fig:elasto_dynamic_u} and~\ref{Fig:elasto_dynamic_v} trace the plane-wave propagation of $u^n_x$ and $u^n_y$ at $n = 99, 106, 149$. The $x$-wave propagates horizontally leftward; the $y$-wave propagates vertically downward. By $n = 106$, high-displacement values penetrate from $\Omega_I$ into $\Omega_{II}$ across the interface $\Gamma_{II}$, and by $n = 149$, the wave fills $\Omega_{II}$ with no visible mismatch at the domain boundary - confirming seamless spatiotemporal coupling via Algorithm~\ref{algo_1}. At $n = 149$, the absolute errors $|u^{149}_{x,\mathrm{FE}} - u^{149}_{x,\mathrm{FE\text{-}FE}}|$ and $|u^{149}_{x,\mathrm{FE}} - u^{149}_{x,\mathrm{FE\text{-}NO}}|$ are $\mathcal{O}(10^{-5})$ and $\mathcal{O}(10^{-4})$, respectively, both negligible relative to displacement magnitudes of $\mathcal{O}(10^{-1})$. As in Case~1, the FE-FE error distribution is asymmetric despite symmetric loading, attributable to the time-dependent evolution of elasto-dynamic fields. Interface-proximal errors are highest for both frameworks, consistent with the inherent challenges of sharp-gradient approximation in scientific machine learning~\cite{wang2024causality}.

\begin{figure}[H]
\begin{center}
\includegraphics[width=1\textwidth]{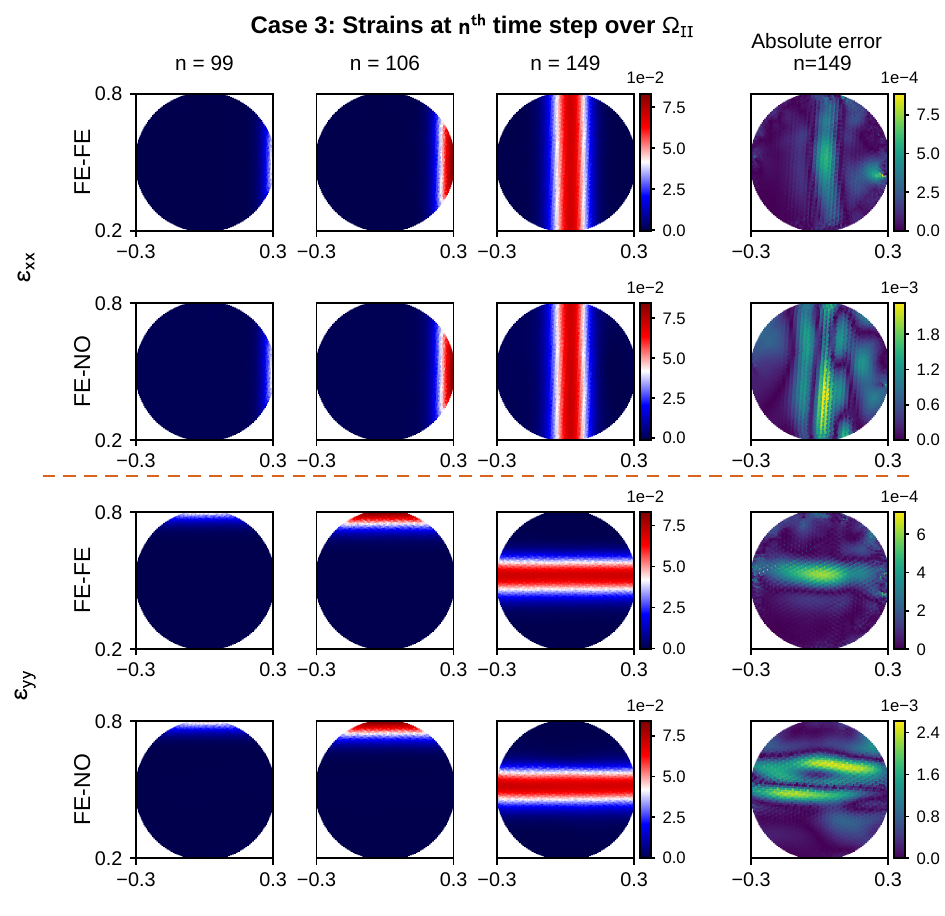}
\caption{Case~3 (Disk Example): Strain evolution in $\Omega_{\mathrm{II}}$ (disk) at
           $n = 99, 106, 149$.
           Rows~1-2: $\varepsilon_{xx}$ from FE-FE and FE-NO;
           rows~3-4: $\varepsilon_{yy}$.
           Column~4: absolute error at $n = 149$.
           A vertical band ($\varepsilon_{xx}$) and a horizontal band
           ($\varepsilon_{yy}$) propagate leftward and downward, consistent
           with Eq.~\eqref{eq:strain_linear}.
           Relative errors remain below $2\%$ (FE-FE) and $5\%$ (FE-NO)
           at all time steps.}
\label{Fig:elasto_dynamic_strain}
\end{center}
\end{figure}

\paragraph{Strain fields}
Figure~\ref{Fig:elasto_dynamic_strain} presents $\varepsilon_{xx}$ and
$\varepsilon_{yy}$ - the first spatial derivatives of $u_x$ and $u_y$ -
at the same three time steps. Both components exhibit a propagating band
structures (a vertical band for $\varepsilon_{xx}$, a horizontal band for
$\varepsilon_{yy}$) that introduce sharp internal interfaces and pose
challenges for both NO and FE solvers. \WW{Peak absolute errors for FE-NO
are $\mathcal{O}(10^{-3}\text{-}10^{-2})$, consistently one order of
magnitude larger than the corresponding displacement errors. This high error originates from (1) differentiation and (2) the difficulty of fixed trunk bias in capturing sharp features, as reported in \cite{lanthaler2022error, kovachki2021universal}.} Relative errors remain below~$5\%$ for FE-NO and
$2\%$ for FE-FE across all time steps, confirming engineering-grade
accuracy.

\begin{figure}[H]
\begin{center}
\includegraphics[width=1\textwidth]{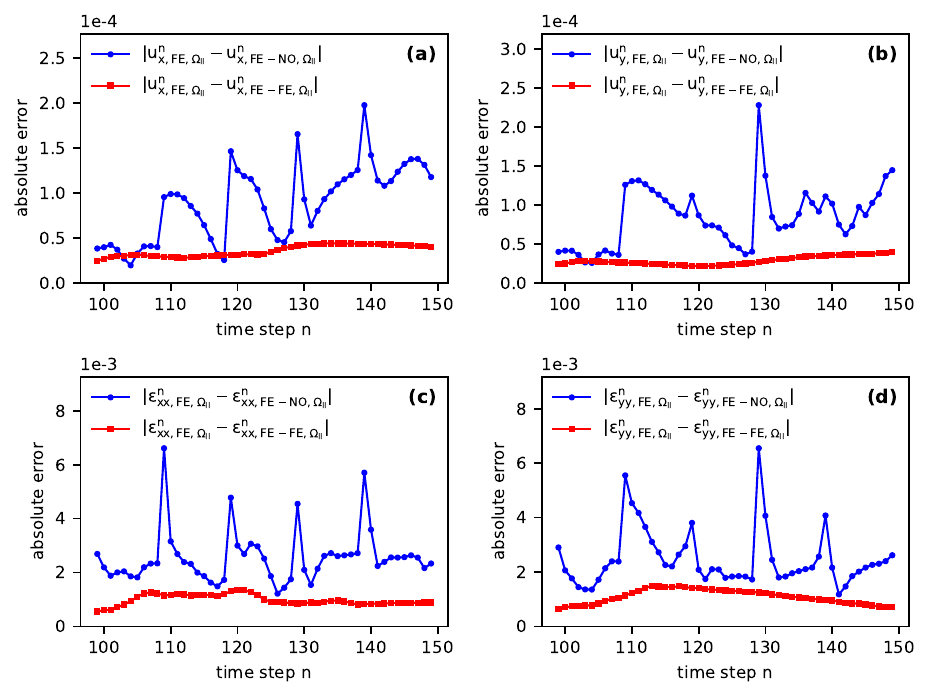}
\caption{Case~3 (Disk Example): Maximum absolute error in $\Omega_{\mathrm{II}}$ (disk) versus time step $n = 99$-$149$ for FE-NO and FE-FE:
           \textbf{(a)}~$|u^n_{x,\mathrm{FE}} - u^n_{x,\mathrm{FE\text{-}NO/FE\text{-}FE}}|$;
           \textbf{(b)}~same for $u_y$;
           \textbf{(c,d)}~same for $\varepsilon_{xx}$ and $\varepsilon_{yy}$.
           Spikes in FE-NO at $n = 110, 120, \ldots$ coincide with transitions
           between pretrained NOs; errors recover within one step, demonstrating
           rapid self-correction.
           Strain errors remain bounded within
           $[10^{-3},\, 7{\times}10^{-3}]$ throughout all 50 steps, preventing
           unbounded auto-regressive error growth.}
\label{Fig:error_dynamic}
\end{center}
\end{figure}

\paragraph{Error accumulation under auto-regression}
Figure~\ref{Fig:error_dynamic} plots the maximum absolute error in
$\Omega_{II}$ against time step $n$ from $n = 99$ to~$149$ for both
coupling schemes. This is a critical diagnostic: auto-regressive methods
can exhibit exponential error growth, as reported in related
work~\cite{wang2025time, michalowska2024neural}. In the present FE-NO framework,
no such growth is observed. FE-FE errors in displacement and strain
vary smoothly within $[0,\,5\times10^{-5}]$ and $[0,\,2\times10^{-3}]$,
respectively. FE-NO errors are more variable, with periodic spike-like
increases at time steps that are integer multiples of~10 ($n = 110, 120,
\ldots$), coinciding with transitions to a new pretrained NO. Each spike
is followed by a sharp decline, demonstrating that the coupling framework
recovers rapidly after each NO switch. Crucially, the strain error
$|\varepsilon_{\mathrm{FE}} - \varepsilon_{\mathrm{FE\text{-}NO}}|$
remains bounded within $[10^{-3},\,7\times10^{-3}]$ throughout all~50
time steps - the absence of accumulation here is particularly important
because strain-derived traction at $\Gamma^{\mathrm{in}}_I$ is the sole
Neumann input to the FE solver in $\Omega_I$. Bounded strain error, therefore, guarantees bounded input data to FE, which in turn returns
reliable displacement values to feed the next NO step, closing the
auto-regressive loop stably.

\begin{figure}[H]
\begin{center}
\includegraphics[width=1\textwidth, height= 8.5cm]{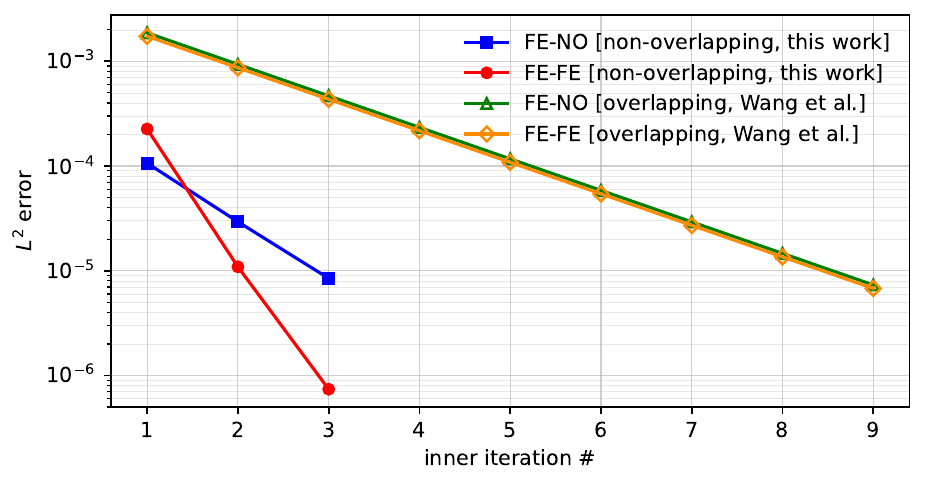}
\caption{Case~3 (regular example): $L^2$ error norm versus inner Schwarz iteration at $n = 139$ in current and prior work \cite{wang2025time}. In this non-overlapping work, both schemes reach $\approx 10^{-4}$ after one inter-domain exchange and converge in 3 total iterations ($\varepsilon = 10^{-5}$); Wall-clock time: FE-NO $1.8\,\mathrm{s}$  and FE-FE $1.5 \,\mathrm{s}$. While in prior overlapping work, both schemes require 9 iterations; wall-clock time: FE-NO $188.8\,\mathrm{s}$ and FE-FE $15.5 \,\mathrm{s}$.}
\label{Fig:Elasto_dynamic_error_profile}
\end{center}
\end{figure}

\paragraph{Schwarz convergence}
Figure~\ref{Fig:Elasto_dynamic_error_profile} shows the $L^2$ error norm against inner iteration count at $n = 139$ for both overlapping and non-overlapping approaches. In this non-overlapping work, the error drops to approximately~$10^{-4}$ for both schemes after a single inter-domain exchange. At the second iteration, FE-FE is found slightly above $10^{-5}$, which needs an additional iteration to converge. While the FE-NO coupling is marginally below $10^{-5}$ after 3 inner iterations.  With the multiple time steps in the elastodynamic example, a small noise can change the inner iteration number in FE-FE and FE-NO,  from 2 to 3 and from 3 to 4, respectively, as noted in Table \ref{tab:comparison}.  The total inner iteration count of ~$\mathbf{3}$ for FE-NO represents a dramatic reduction from the ~$\mathbf{9}$ iterations reported for \emph{both} FE-FE and FE-NO in
our prior overlapping Dirichlet-Dirichlet framework~\cite{wang2025time}.
This three-fold reduction confirms that the non-overlapping Neumann-Dirichlet interface coupling is more efficient, explaining the speed up of FE-FE from $15.5 \,\mathrm{s}$ to $1.5 \,\mathrm{s}$. The significant difference in FE-NO, $188.8\,\mathrm{s}$ in overlapping and $1.8\,\mathrm{s}$ in non-overlapping schemes, originates from the architecture of Branch 2. In our prior work, the employed CNN in branch 2 demands uniform mesh nodes over $\Omega_{\mathrm{II}}$ as input, requiring a time consuming interpolation at  each iteration. The PointNet breaks this restriction and leads to a comparable simulation efficiency between FE-FE and FE-NO in elastodynamics.
 
\subsubsection{Case 4: Irregular inner domain (filleted L-shape)}
\label{sec:dynamic_lshape}
 
The L-shaped inner domain is formed by two orthogonal rectangular arms
(horizontal: $1\times0.5$ units; vertical: $0.5\times1$ units) with
fillet radius~$0.1$ at each interior corner. A single pretrained
NO covers all~80 dynamic time steps; no NO switching occurs, so the
spike-like errors of Section~\ref{sec:dynamic_disk} are absent.
 
This geometry cannot be handled by the CNN Branch net
of~\cite{wang2025time}, which requires a regular grid and thus restricts $\Omega_{II}$ to a square. The PointNet based Branch network introduced in this work directly ingests the unstructured FE mesh of the L-shaped domain, extending the time-marching DeepONet to irregular geometries without any preprocessing.

\begin{figure}[H]
\begin{center}
\includegraphics[width=1\textwidth]{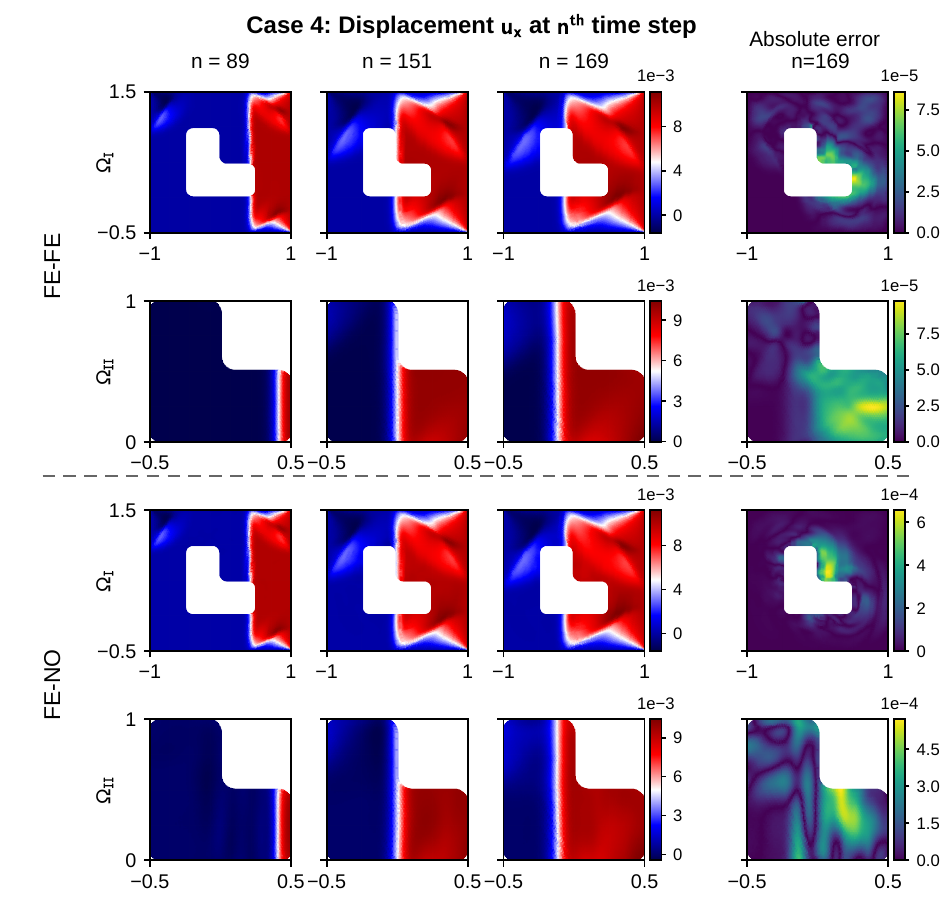}
\caption{Case~4 (L-shaped Example): $u_x$ wave propagation at $n = 89, 151, 169$. The horizontal arm of $\Omega_{\mathrm{II}}$ confines $u_x$
           propagation; the wavefront reaches the arm terminus by $n = 151$.
           Absolute errors at $n = 169$: $\mathcal{O}(10^{-4})$ for both
           schemes, confirming robustness to domain
           irregularity.
           Unlike the overlapping framework of~\cite{wang2025time}, which
           restricts $\Omega_{\mathrm{II}}$ to a square, the PointNet Branch net
           handles this non-convex geometry without preprocessing.}
\label{Fig:L_shape_u}
\end{center}
\end{figure}

\begin{figure}[H]
\begin{center}
\includegraphics[width=1\textwidth]{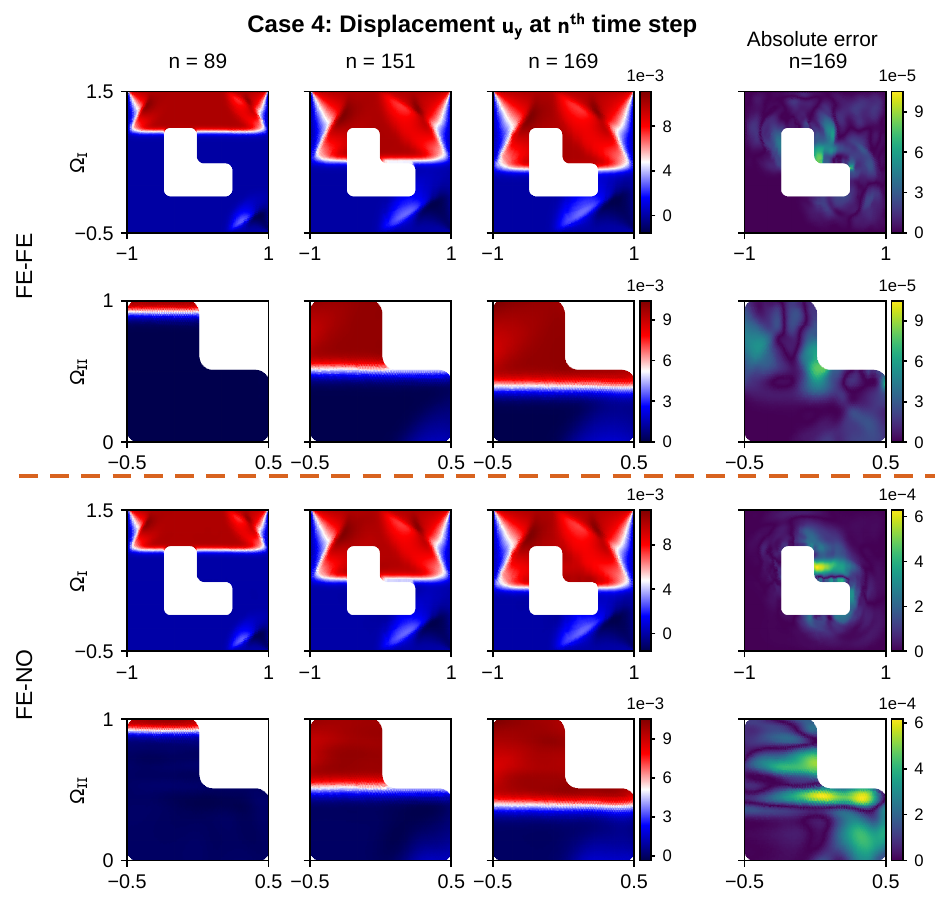}
\caption{Case~4 (L-shaped Example): As Figure~\ref{Fig:L_shape_u} but for $u_y$ (L-shaped case). The vertical arm governs $u_y$ propagation.
           Maximum absolute errors for FE-NO reach ${\approx}\,6{\times}10^{-4}$
           in both domains, negligible relative to field magnitudes of
           $\mathcal{O}(10^{-1})$.}
\label{Fig:L_shape_v}
\end{center}
\end{figure}
 
\paragraph{Displacement fields and errors}
Figures~\ref{Fig:L_shape_u} and~\ref{Fig:L_shape_v} show the wave
propagation at $n = 89, 151, 169$. At $n = 89$, the high-displacement
wavefront enters $\Omega_{II}$ through the top and right boundaries of
the L-shape, after which $u_x$ propagates exclusively within the
horizontal arm and $u_y$ within the vertical arm - a direct consequence
of the domain topology. By $n = 151$ the waves reach the respective arm
termini, and by $n = 169$ the fields continue to evolve smoothly.
 
Absolute errors at $n = 169$ are $\mathcal{O}(10^{-4})$ for FE-NO in
both domains and $\mathcal{O}(10^{-4})$ for FE-FE (with the maximum
near the right edge of the L-shape). These levels are comparable to
those of the regular disk case ($10^{-5}$ for FE-FE and $10^{-4}$ for
FE-NO), confirming that the coupling scheme is robust to domain
irregularity.

\begin{figure}[H]
\begin{center}
\includegraphics[width=1\textwidth]{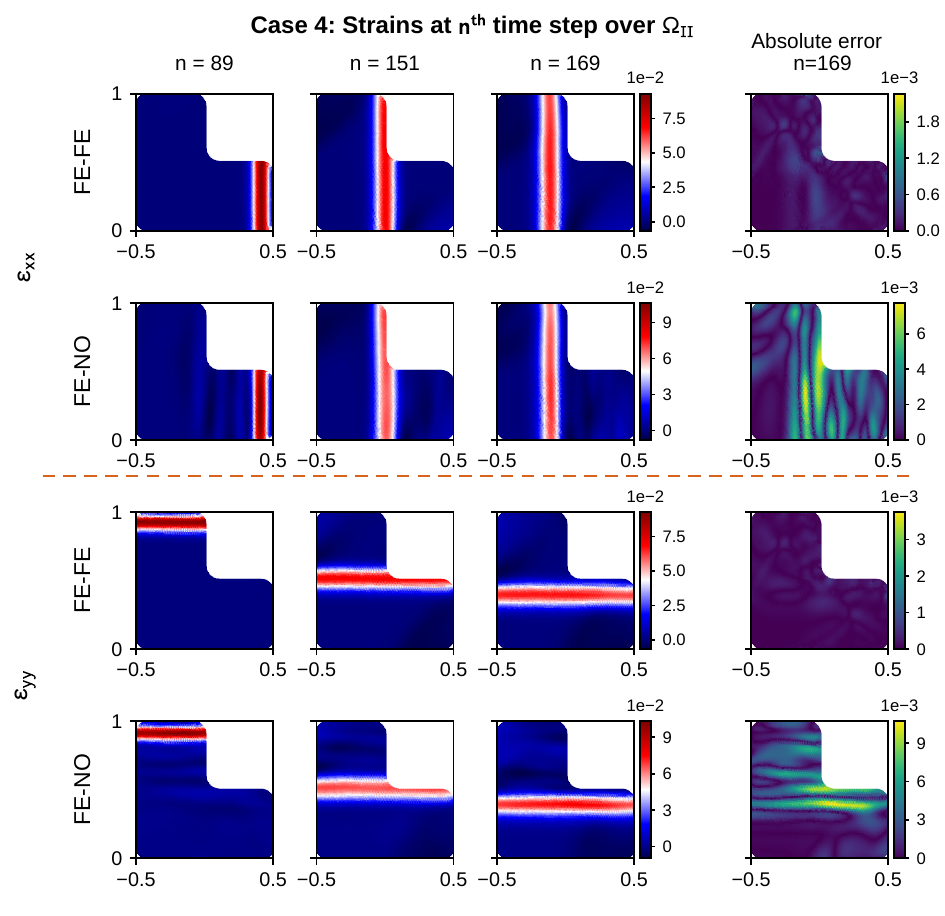}
\caption{Case~4 (L-shaped Example): Strain evolution in $\Omega_{\mathrm{II}}$ at $n = 89, 151, 169$. Peak errors concentrate at the transitions between straight segments and quarter-circle fillets, where curvature changes abruptly.
           FE-FE relative strain errors are ${<}\,4\%$; FE-NO maximum
           relative error reaches ${\approx}\,20\%$ at $n = 89$ and decreases
           rapidly thereafter as alternating stripe artefacts dissipate by
           $n = 120$.}
\label{Fig:L_shape_strain}
\end{center}
\end{figure}

\paragraph{Strain fields}
Figure~\ref{Fig:L_shape_strain} shows $\varepsilon_{xx}$ and
$\varepsilon_{yy}$ at the same three time steps. As in the disk case,
propagating band structures are observed, and the highest errors are
concentrated at curvature-discontinuity points - the transitions between
the straight segments and the quarter-circle fillets. The FE-FE strain
errors ($\mathcal{O}(10^{-3})$, relative error $<4\%$) are higher and
more variable than in the disk case, owing to the geometric
singularities at the corners. For FE-NO, the maximum absolute errors
reach approximately~$8\times10^{-2}$ for $\varepsilon_{xx}$ and
$10^{-2}$ for $\varepsilon_{yy}$, with the highest values at $n = 89$
(the first step of NO inference); relative errors exceed~$10\%$
at that step. Alternating stripe patterns in
$\varepsilon^{89}_{xx,\mathrm{FE\text{-}NO}}$ and
$\varepsilon^{89}_{yy,\mathrm{FE\text{-}NO}}$ dissipate by $n = 151$,
demonstrating the self-correcting tendency of the coupled framework.
These elevated early-step errors indicate that the current NO
generalization across large geometric curvature changes remains limited,
and suggest that online parameter updating or transfer learning - areas
identified for future work - could bring the L-shaped case to the same
accuracy level as the disk.

\begin{figure}[H]
\begin{center}
\includegraphics[width=1\textwidth]{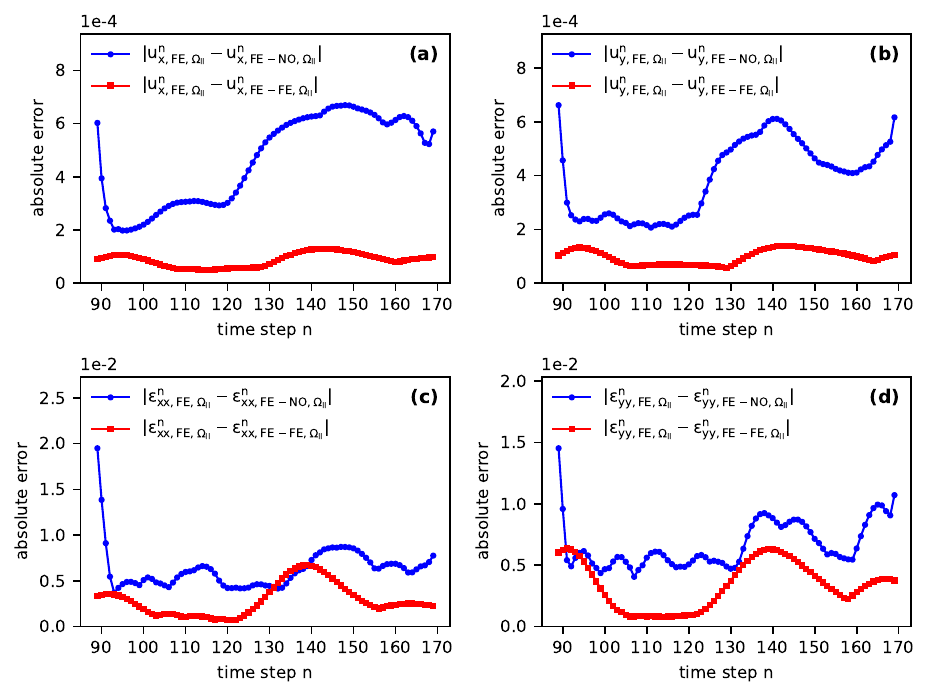}
\caption{Case~4 (L-shaped Example): Maximum absolute error in $\Omega_{\mathrm{II}}$ versus time step $n = 89$-$169$. A single NO governs all 80 steps; the periodic spikes of Figure~\ref{Fig:error_dynamic} are absent and errors evolve smoothly.
           Despite an overall increasing trend in FE-NO errors, the Schwarz
           iteration converges within 3 inner iterations at every time step.
           The sharp decline over the first five steps reflects the stabilising
           mechanism described in Section~\ref{sec:dynamic_disk}: bounded
           strain-derived traction at $\Gamma^{\mathrm{in}}_{\mathrm{I}}$
           prevents unbounded error propagation in the auto-regressive loop.}
\label{Fig:error_L_shape}
\end{center}
\end{figure}
 
\paragraph{Error accumulation}
Figure~\ref{Fig:error_L_shape} plots the maximum absolute error against
time step $n = 89$ to~$169$. Because a single NO governs all~80 steps,
the periodic spikes of the disk case are absent: FE-NO errors evolve
smoothly, albeit with an overall increasing trend for both displacement
and strain. Despite these larger errors, the Schwarz iteration converges
within~3 inner iterations at every time step, confirming the robustness
of the non-overlapping coupling scheme. The initial~5 time steps show a
sharp decline in $|u^n_{\mathrm{FE},\Omega_{II}} -
u^n_{\mathrm{FE\text{-}NO},\Omega_{II}}|$ from the high starting values,
after which errors grow at a moderate pace. This behaviour underscores
the same stabilising mechanism identified in the disk case: bounded
strain inputs to FE prevent unbounded error propagation in the
auto-regressive loop.

\subsection{Case 5: Static 3D (Quarter-tube)}
\label{3D_tube}

\begin{figure}[H]
\begin{center}
\includegraphics[width=1\textwidth]{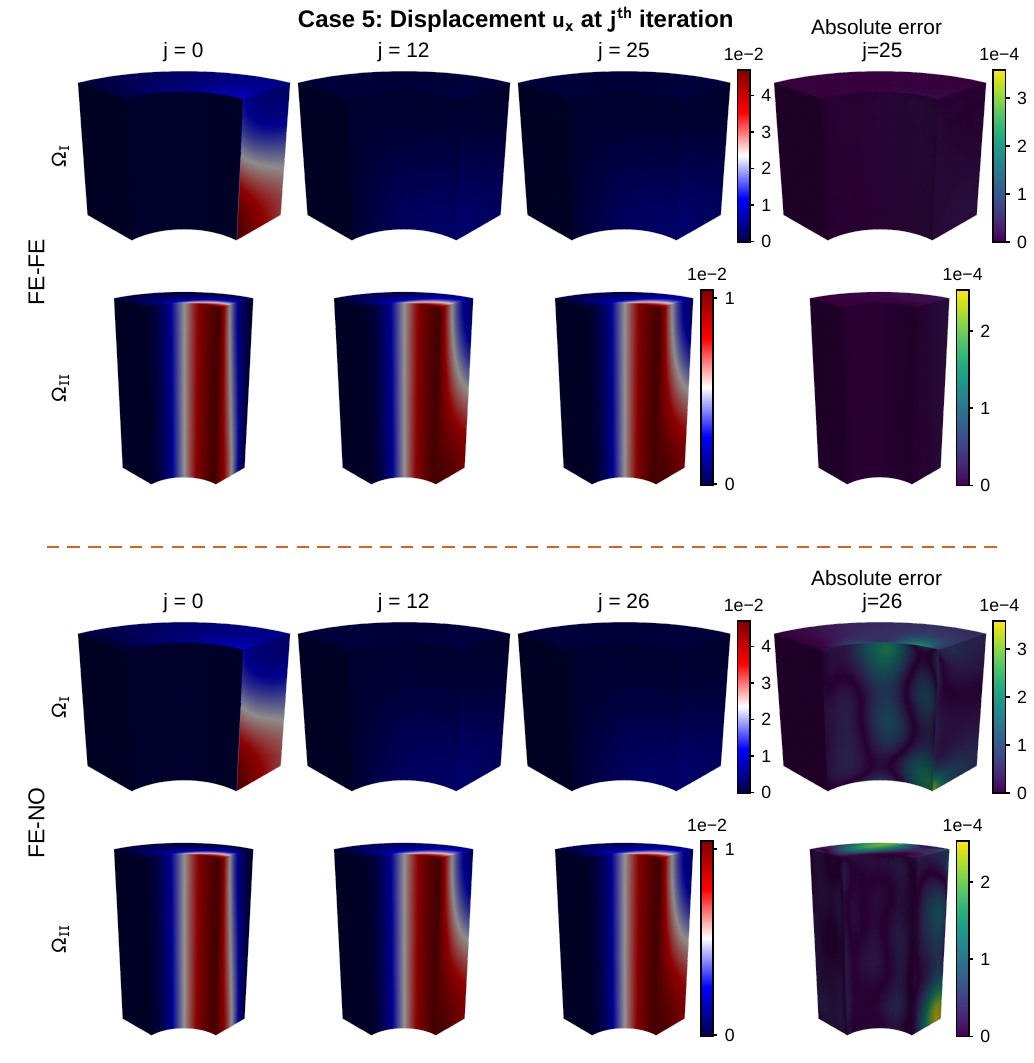}
\caption{\WW{Case~5: Schwarz iteration convergence of $u_x$ for the static linear-elastic 3D quarter-tube. FE-FE results are shown in the top two rows and FE-NO results in the bottom two rows for selected Schwarz iterations. The fourth column shows the absolute error $|u_x^{\mathrm{ref}}-u_x^{j_{\mathrm{cv}}}|$ at convergence. FE-NO reaches convergence after 26 iterations, compared with 25 iterations for FE-FE. The maximum relative error at convergence is below $3\%$ for both approaches; the largest FE-NO error occurs near the right-bottom corner of the interface due to the local singularity.}}
\label{Fig:static_u_3d}
\end{center}
\end{figure}

\begin{figure}[H]
\begin{center}
\includegraphics[width=1\textwidth]{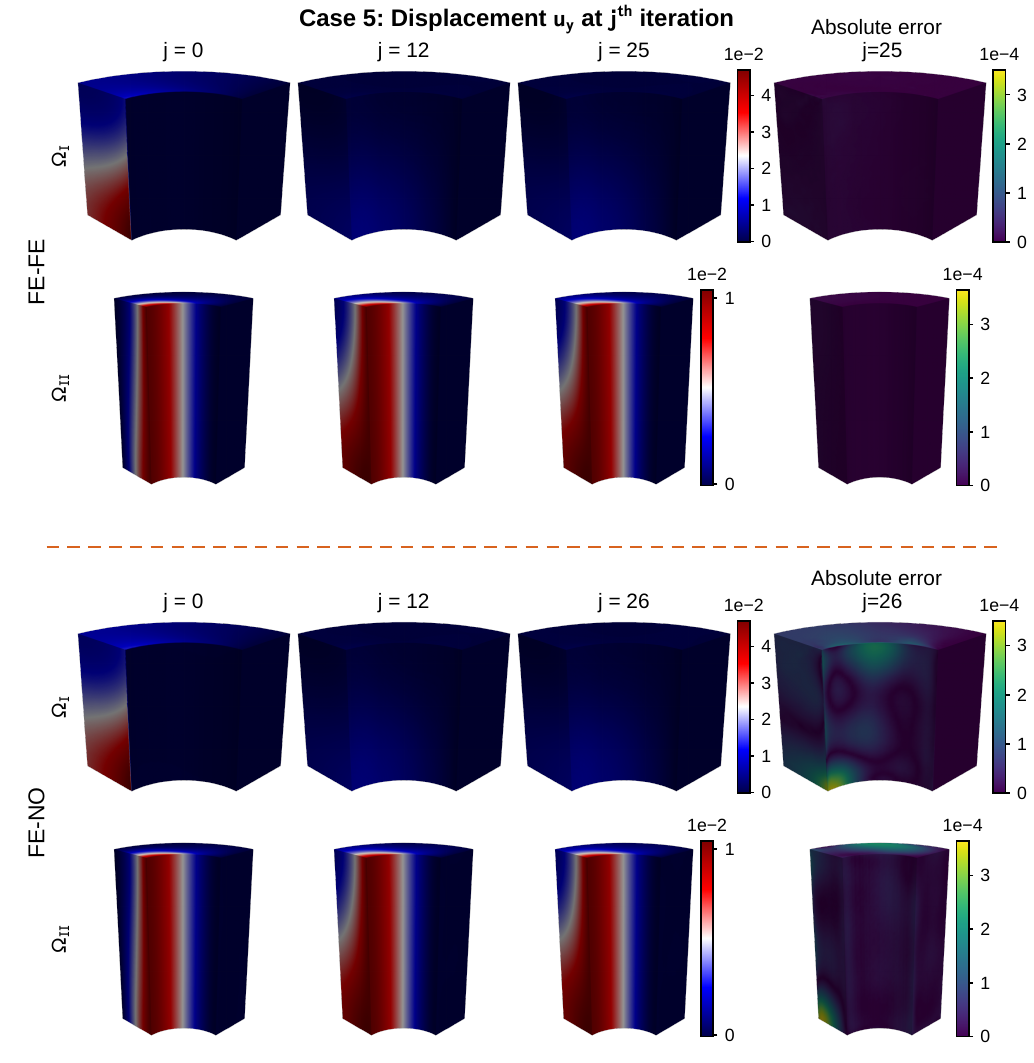}
\caption{\WW{Case~5: As Figure~\ref{Fig:static_u_3d}, but for $u_y$. FE-NO converges after 26 Schwarz iterations, compared with 25 iterations for FE-FE. The maximum relative error at convergence is below $3.5\%$ for both approaches, with the largest error occurring near the left-bottom corner of the interface.}}
\label{Fig:static_v_3d}
\end{center}
\end{figure}

\begin{figure}[H]
\begin{center}
\includegraphics[width=1\textwidth]{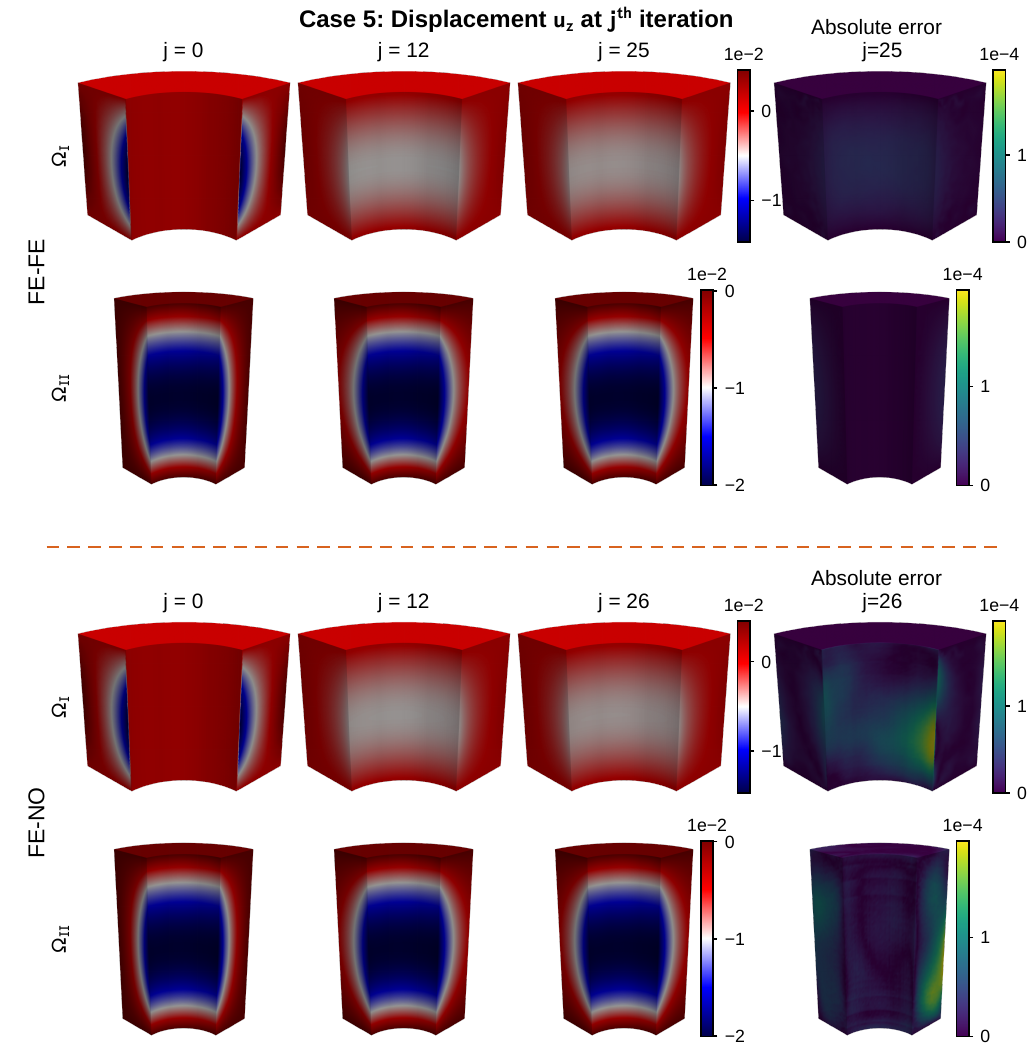}
\caption{\WW{Case~5: As Figure~\ref{Fig:static_u_3d}, but for $u_z$. The maximum relative error at convergence is below $2\%$ for both approaches.}}
\label{Fig:static_w_3d}
\end{center}
\end{figure}

\WW{A parabolic static displacement loading $\mathbf{u} = (u_x, u_y, u_z) = (0.01\times x^2, 0.01 \times y^2, 0.005\times z(z-4))$ is prescribed on the inner cylindrical surface. The left, right, top, and bottom boundaries are subjected to roller boundary conditions, as illustrated in Figure \ref{fig:Schematics_3D_tube}. The reference solutions for $u_x^{\mathrm{ref}}$, $u_y^{\mathrm{ref}}$, and $u_z^{\mathrm{ref}}$ are obtained from a standalone FE simulation on the intact domain and are used to evaluate the errors reported in Figures \ref{Fig:static_u_3d}-\ref{Fig:static_w_3d}. The governing equations are identical to Eqs. \ref{eq:static_res_x}-\ref{eq:static_bc_eps}, with the only difference being an additional $z$-dimension: }
\begin{align}
  &(\lambda + 2\mu)\frac{\partial^2 \mathcal{G}^{u_x}_{\bm\theta_1}}{\partial x^2}
  + \mu\left(\frac{\partial^2 \mathcal{G}^{u_x}_{\bm\theta_1}}{\partial y^2} + \frac{\partial^2 \mathcal{G}^{u_x}_{\bm\theta_1}}{\partial z^2} \right)
  + (\lambda + \mu)\left(\frac{\partial^2 \mathcal{G}^{u_y}_{\bm\theta_2}}{\partial x\,\partial y} + \frac{\partial^2 \mathcal{G}^{u_z}_{\bm\theta_3}}{\partial x\,\partial z} \right)
  = 0, \quad (x, y, z)\in\Omega_{II}, \label{eq:static_res_x_3D} \\
   & (\lambda + 2\mu)\frac{\partial^2 \mathcal{G}^{u_y}_{\bm\theta_2}}{\partial y^2}
  + \mu\left(\frac{\partial^2 \mathcal{G}^{u_y}_{\bm\theta_2}}{\partial x^2} + \frac{\partial^2 \mathcal{G}^{u_y}_{\bm\theta_2}}{\partial z^2} \right)
  + (\lambda + \mu)\left(\frac{\partial^2 \mathcal{G}^{u_x}_{\bm\theta_1}}{\partial x\,\partial y} + \frac{\partial^2 \mathcal{G}^{u_z}_{\bm\theta_3}}{\partial y\,\partial z} \right)
  = 0, \quad (x, y, z)\in\Omega_{II}, \label{eq:static_res_y_3D} \\
     & (\lambda + 2\mu)\frac{\partial^2 \mathcal{G}^{u_z}_{\bm\theta_3}}{\partial z^2}
  + \mu\left(\frac{\partial^2 \mathcal{G}^{u_z}_{\bm\theta_3}}{\partial x^2} + \frac{\partial^2 \mathcal{G}^{u_z}_{\bm\theta_1}}{\partial y^2} \right)
  + (\lambda + \mu)\left(\frac{\partial^2 \mathcal{G}^{u_x}_{\bm\theta_1}}{\partial x\,\partial z} + \frac{\partial^2 \mathcal{G}^{u_y}_{\bm\theta_2}}{\partial y\,\partial z} \right)
  = 0, \quad (x, y, z)\in\Omega_{II}, \label{eq:static_res_z_3D} \\
      &\mathcal{G}^{u_x}_{\bm\theta_1} = u_x(x,y,z), \quad
  \mathcal{G}^{u_y}_{\bm\theta_2} = u_y(x,y,z), \quad \mathcal{G}^{u_z}_{\bm\theta_3} = u_z(x,y,z),  \quad (x,y,z)\in\Gamma_{II},
  \label{eq:static_bc_u_3D}\\
   & \mathcal{G}^{\varepsilon_{xx}}_{\bm\theta_1,\bm\theta_2} = \varepsilon_{xx},\quad
  \mathcal{G}^{\varepsilon_{xy}}_{\bm\theta_1,\bm\theta_2} = \varepsilon_{xy},\quad
   \mathcal{G}^{\varepsilon_{xz}}_{\bm\theta_1,\bm\theta_3} = \varepsilon_{xz}, \quad \nonumber\\
  &\mathcal{G}^{\varepsilon_{yy}}_{\bm\theta_1,\bm\theta_2} = \varepsilon_{yy}, \quad
   \mathcal{G}^{\varepsilon_{yz}}_{\bm\theta_2,\bm\theta_3} = \varepsilon_{yz}, \quad
    \mathcal{G}^{\varepsilon_{zz}}_{\bm\theta_3,\bm\theta_3} = \varepsilon_{zz},
  \quad (x,y,z)\in\Gamma_{II},
  \label{eq:static_bc_eps_3D}
\end{align}
\WW{where $\mathcal{G}^{u_z}_{\bm\theta_3}$ denotes the $z$-component displacement operator. The displacement field prescribed at the FE--NO interface is generated using the same Gaussian random field (GRF) procedure described in \ref{Data_generation}. In the present 3D problem, the sampled displacement fields are prescribed only on the cylindrical interface $\Gamma^{r=2}_{II}$ and serve as the input to Branch 1 of the neural operator. The resulting test loss, evaluated on five unseen samples, decreases to approximately $3\times10^{-4}$ after $10^6$ training epochs. This demonstrates that the pretrained NO can accurately represent the local three-dimensional solution operator and provides the basis for the subsequent FE--NO coupling experiment.}

\begin{figure}[H]
\begin{center}
\includegraphics[width=1\textwidth]{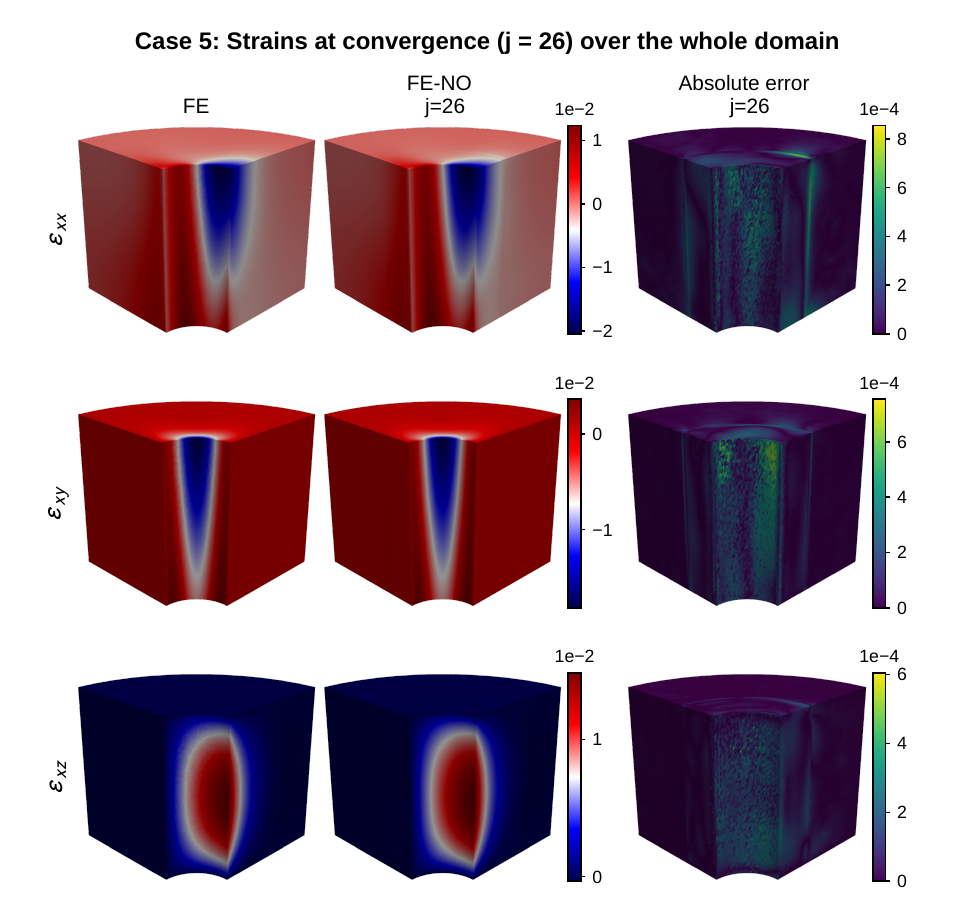}
\caption{\WW{Case~5: Converged strain fields from FE-NO at j=26, compared with the reference FE solution. The first, second, and third columns show the reference solution, FE-NO prediction, and absolute error, respectively. The strain components are obtained analytically from the predicted displacement field. The maximum relative strain error is below $8\%$}}
\label{Fig:strains1_3d}
\end{center}
\end{figure}

\begin{figure}[H]
\begin{center}
\includegraphics[width=1\textwidth]{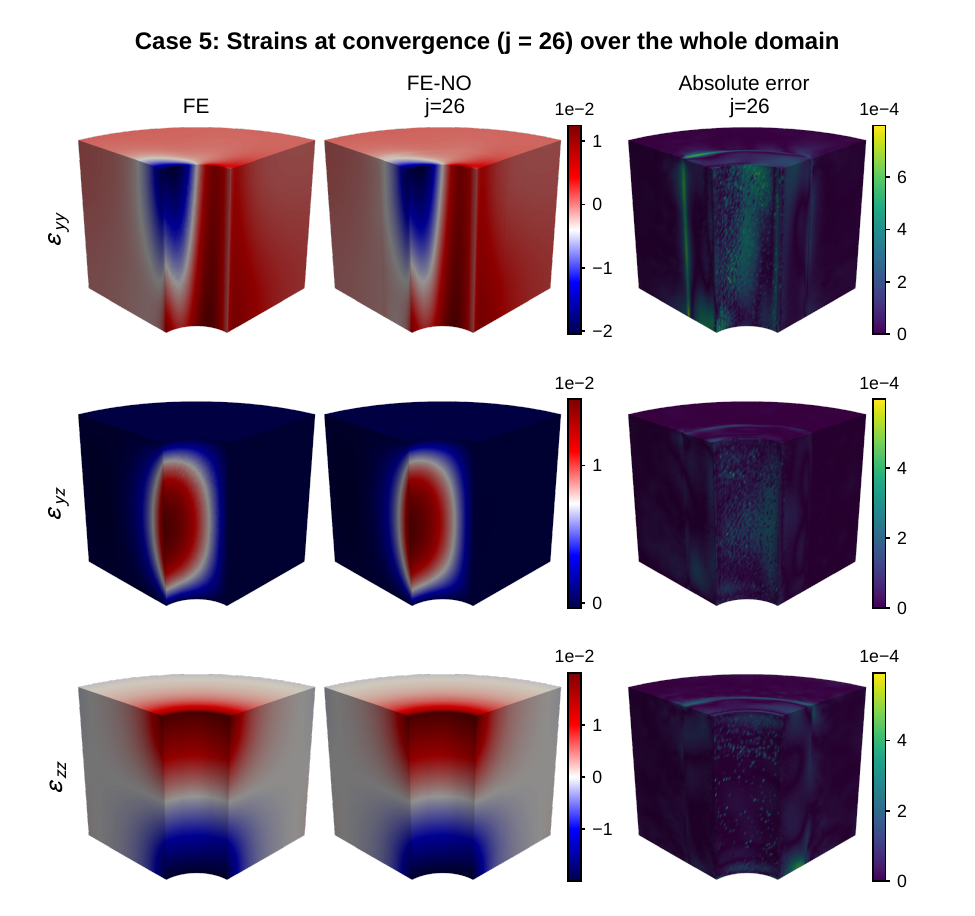}
\caption{\WW{Case~5: As Figure \ref{Fig:strains1_3d}  but for $\varepsilon_{yy}$, $\varepsilon_{yz}$, and $\varepsilon_{zz}$; the maximum relative strain error is $8\%$.}}
\label{Fig:strains2_3d}
\end{center}
\end{figure}

\begin{figure}[H]
\begin{center}
\includegraphics[width=0.75\textwidth]{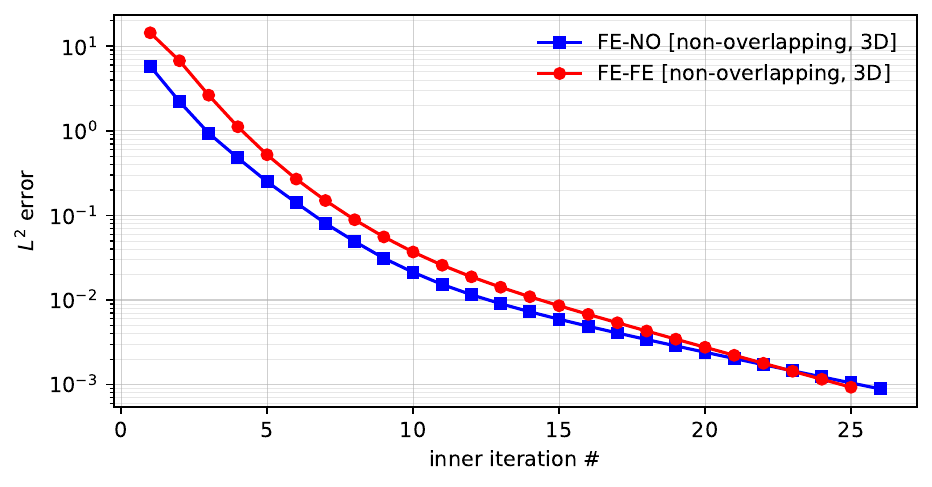}
\caption{\WW{Case~5: $L^2$ error norm (Eq.~\eqref{eq:convergence}) versus inner Schwarz iteration in our current 3D quarter-tube, FE-NO converges in 26 iterations; FE-FE requires 25. Wall-clock times: FE-NO $175.8\,\mathrm{s}$, FE-FE $1,014.1\,\mathrm{s}$ ($5.76{\times}$ faster).}}
\label{Fig:static_error_3d}
\end{center}
\end{figure}

\begin{figure}[H]
\begin{center}
\includegraphics[width=0.6\textwidth]{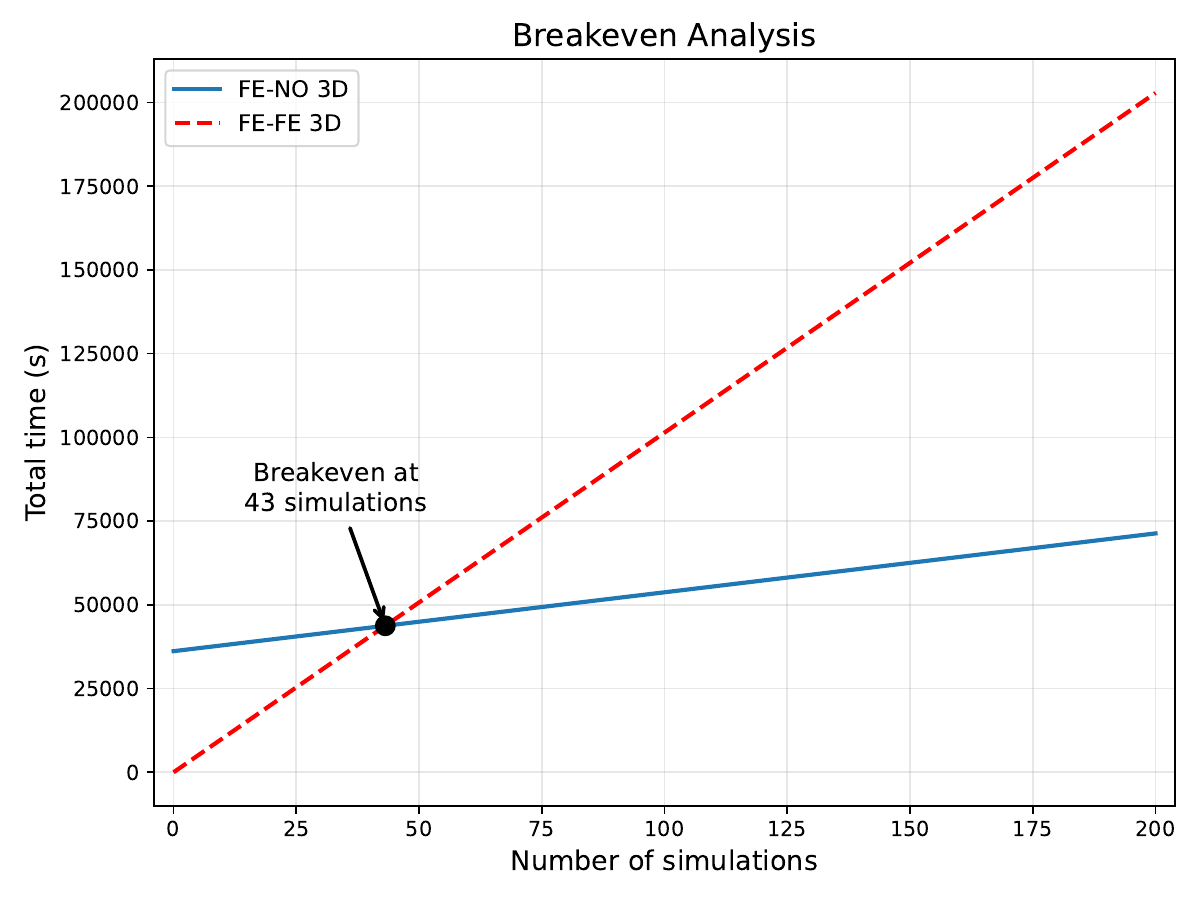}
\caption{\WW{Case~5: Cumulative computational cost of FE-NO and FE-FE as a function of the number of forward simulations. FE-NO incurs a one-time offline cost of 36,157 s for data generation and training, followed by 175.8 s per forward simulation, whereas FE-FE requires 1,014.1 s per simulation. The cumulative costs become equal after approximately 43 simulations.}}
\label{Fig:comput_time}
\end{center}
\end{figure}

\subsubsection{Coupling results}

\WW{The evolution of the displacement components $u_x^j$, $u_y^j$, and $u_z^j$ during the inner Schwarz iterations is shown in Figures~\ref{Fig:static_u_3d}-\ref{Fig:static_w_3d} for both FE-FE and FE-NO coupling. At the initial iteration ($j=0$), discontinuities are present across the FE--FE and FE--NO interface $\Gamma^{r=2}$, as the two subdomain solutions are initialized independently. The displacement fields also exhibit strong concentration near the loaded inner cylindrical surface $\Gamma^{r=1}_{II}$. After approximately 12 Schwarz iterations, the interface discontinuity is substantially reduced and the displacement fields become continuous across $\Gamma^{r=2}$. Further Schwarz iterations progressively reduce the interface mismatch until the prescribed convergence criterion, $L^2$ error $<10^{-3}$, is satisfied. FE-NO requires 26 iterations, compared with 25 iterations for FE-FE, demonstrating that replacing the high-resolution FE subdomain by the pretrained NO does not materially degrade the convergence of the Schwarz coupling.}

\WW{At convergence, the FE-NO displacement fields agree closely with the reference full-domain FE solution. The maximum relative errors are below $3\%$, $3.5\%$, and $2\%$ for $u_x$, $u_y$, and $u_z$, respectively. The largest errors are localized primarily near interface corners, where the solution exhibits a local singularity. The corresponding strain fields, shown in Figures~\ref{Fig:strains1_3d}-\ref{Fig:strains2_3d}, also agree well with the reference FE solution, with maximum relative errors below $8\%$. These results demonstrate that the NO can replace the high-resolution FE subdomain while maintaining sufficient accuracy in both displacement and strain fields.}

\paragraph{Convergence efficiency and computational speed}

\WW{The 3D quarter-tube provides a particularly relevant test of the computational advantage of the proposed FE-NO framework because both subdomains contain more than $100,000$ degrees of freedom. The FE solver therefore employs a conjugate-gradient (CG) preconditioner~\cite{petsc-web-page,stiefel1952methods}, and each Schwarz iteration requires an additional iterative FE solve in the FE-FE formulation. In contrast, the NO subdomain is evaluated by a forward pass through the pretrained neural operator, whose inference time is negligible compared with the corresponding FE solve. FE-NO converges in 26 Schwarz iterations with a total wall-clock time of $175.8,\mathrm{s}$, whereas FE-FE converges in 25 iterations and requires $1,014.1,\mathrm{s}$. Thus, FE-NO is approximately $5.76\times$ faster than FE-FE for a single forward simulation. This result directly demonstrates the intended role of the neural operator in the proposed framework: it does not replace the FE solver over the entire domain, but instead removes the repeated computational cost associated with solving the expensive high-resolution subdomain. As shown in Figure~\ref{Fig:comput_time}, the cumulative cost of FE-NO becomes lower than that of FE-FE after approximately 43 forward simulations. This illustrates an important characteristic of the proposed framework: the offline training cost is amortized over repeated simulations, making the approach particularly attractive for applications requiring many forward solves with similar localized physics.}

\section{Conclusion}
\label{sec:summary}
This work develops a hybrid FE-NO framework that exploits the complementary strengths of both: FE governs the coarse, well-behaved outer domain; a pretrained PI-DeepONet governs the computationally intensive inner subdomain; and a non-overlapping Schwarz alternating method with Neumann-Dirichlet interface exchange provides the coupling mechanism. Three concrete advances over our previous overlapping FE-NO framework~\cite{wang2025time} are demonstrated. First, stress and strain operators are derived analytically from the displacement operators through the kinematic equations, rather than trained as independent networks. This reduces the number of trainable parameter sets from five~\cite{wang2025time} to two while enforcing mechanical consistency by construction; the resulting maximum relative displacement error remains below 0.5\% in static and quasi-static cases and 2.5\% in dynamic simulations. Second, a PointNet is embedded in the time-marching branch network, enabling the NO to ingest kinetic fields directly from unstructured FE meshes without interpolation. This extends the framework from the regular-grid subdomains of \cite{wang2025time} to arbitrarily shaped geometries, as demonstrated on a filleted L-shaped subdomain. Third, the non-overlapping Neumann-Dirichlet interface formulation reduces the number of inner Schwarz iterations from nine in \cite{wang2025time} to three per time step in elasto-dynamic simulations, while maintaining bounded error accumulation over all tested time horizons (up to 80 steps). In the static and quasi-static cases, the FE-NO solver is $2.7\times$ and $1.75\times$ faster than the equivalent FE-FE solver, respectively. \WW{The 3D quarter-tube example further demonstrates the scalability of the framework, with FE-NO achieving a $5.77\times$ reduction in computational cost relative to FE-FE while maintaining displacement errors below 3.5\%.}

\paragraph{Future work} Three directions are identified. First, the PointNet architecture naturally extends to three-dimensional unstructured meshes; \WW{already validated on static example in case 5,  the Point-DeepONet-FE framework on more complex 3-D solid mechanics problems (e.g., a notched specimen under fatigue loading) is the most immediate next step}. Second, the current subdomain assignment is prescribed a priori; an adaptive decomposition strategy, driven by local error indicators or nonlinearity measures, would eliminate this assumption and make the framework applicable to problems where critical regions are not known in advance. Third, for the L-shaped case and more complex geometries, an online transfer-learning scheme in which the pretrained NO is fine-tuned at each time step using a small number of FE evaluations is expected to reduce early-step strain errors to the same level as the disk case, closing the remaining accuracy gap without a full retraining cycle.

\section*{Author contributions}
\noindent Conceptualization: WW, SG \\
Investigation: WW, AG, SG, HHR\\
Visualization: WW, SG\\
Supervision: SG, HHR\\
Writing - original draft: WW\\
Writing-review \& editing: WW, AG, SG, HHR

\section*{Acknowledgements}
The authors (WW and HHR) would like to acknowledge the support by the Hong Kong General Research Fund under Grant Numbers 15213619 and 15210622, and by an industry collaboration project (HKPolyU Project ID: P0039303). SG is supported by the 2024 Johns Hopkins Discovery Award and National Science Foundation Grant Number 2438193. The authors acknowledge the computational resources provided by the University Centre for AI Computing (UCAIC) and Centre for Large AI Models (CLAIM) at The Hong Kong Polytechnic University.

\section*{Data and code availability}
\noindent All codes and datasets will be made publicly available at {\small\url{https://github.com/Centrum-IntelliPhysics/Time-Marching-Non-overlapping-Neural-operator-FE-Coupling} upon publication of the work.

\section*{Competing interests}
\noindent The authors declare no competing interests.

\bibliographystyle{elsarticle-num}  
\bibliography{ref} 

\newpage
\renewcommand{\thetable}{A\arabic{table}}  
\renewcommand{\thefigure}{A\arabic{figure}} 
\makeatother
\setcounter{figure}{0}
\setcounter{table}{0}
\setcounter{section}{1}
\setcounter{page}{0}
\appendix 
 \section{Data generation }
\label{Data_generation}

\WW{In all five solid mechanics examples, both the displacements and strains (or stresses) at the non-overlapping interface $\Gamma_{II}$ are required to train the corresponding NOs. Notably, these boundary displacements and strains are not entirely independent. FE solvers in FEniCSx are consequently used over an intact computational domain to generate these correlated values.} The schematics of all four 2D loading cases are displayed in Figure \ref{fig:Schematics_loss} The first three cases share the same schematics: an intact square with length 2 unit is fragmented by an inner circle, resulting in two subdomains $\Omega_{I}$ and $\Omega_{II}$. The inner circular interface, $\Gamma_{II}$, consists of uniform 100 vertices with a 0.3 unit radius. Due to the implementation of second order continuous Galerkin method, 200 nodes are uniformly distributed at $\Gamma_{II}$ to facilitate the further data collection. 

For both static and quasi-static case, the loading conditions for data generation are similar: the bottom edge of this intact square is fixed with top edge prescribed by a static loading in both $x$ and $y$ directions. In a static regime, 1000 different $x$- and $y$-displacements are generated using a Gaussian Random Field (GRF) with value scale parameter \WW{$s_{u_x} = s_{u_y} = 0.01$} and length scale parameter \WW{$l_{u_x}= l_{u_y} = 0.2$}. Then, these displacements are loaded on the top edge of this linear elastic model to generate 1000 pairs $\mathbf{u}$ and $\boldsymbol{\varepsilon}$ at $\Gamma_{II}$, which are used to train the static NOs in section \ref{sec:static}. Similarly in quasi-static case, 1000 GRF-generated displacements in both directions are applied on top edge of the hyper-elastic model within large deformation framework, with \WW{$s_{u_x}=s_{u_y}=0.1$} and \WW{$l_{u_x}=l_{u_y}=1$}. And the hyper-elastic dataset is obtained by collecting the 1000 corresponding $\mathbf{u}_{|\Gamma_{II}}$ and $\boldsymbol{\varepsilon}_{|\Gamma_{II}}$. 

In elastodynamics, 100 random displacements for $x$ and $y$ directions are generated using GRF. The left and bottom edges of this linear-elastic square are clamped, while the right are top edges are subjected to the newly obtained loading displacements in $x$- and $y$-directions, respectively. It is noteworthy that this model is considered in dynamic regime, and Newmark-$\beta$ time stepping scheme is implemented in FEniCSx. In disk inner domain case, for each 11 time steps, we extract not only the boundary $\mathbf{u}_{|\Gamma_{II}}^{n}$ and $\boldsymbol{\varepsilon}_{|\Gamma_{II}}^{n}$, but also the domain displacement and velocity - $\mathbf{u}^{n-1}$ and $\dot{\mathbf{u}}^{n-1}$ over $\Omega_{II}$. In such dataset, the domain kinetic quantities from 0 to 10 time steps are fed to Branch network, while the boundary values from time step 1 through 11 are encoded as inputs of Branch~1 work, yielding 1000 total training samples (10 time steps $\times$ 100 samples). Similarly, for the filleted L-shaped case, we employ a kinetic dataset comprising 81 time steps-totaling 8,000 samples (80 time steps $\times$ 100 samples)-to train the corresponding NOs. 

\WW{For linear elastic 3D quarter-tube in static regime, a 2D GRF is utilized at cylindrical interface with value scale parameters ($s_{u_x} = s_{u_y} = s_{u_z} = 0.01$) and length scale in both circumferential and axial directions ($\bm{l}_{u_x} = \bm{l}_{u_y} = (2, 2)$ and  $\bm{l}_{u_z} = (6, 7)$). Then, similar to 2D static case, 1000 pairs $\mathbf{u}$ and $\mathbf{\varepsilon}$ are obtained at the interface as the training dataset.}

All PI-DeepONets were trained on a single NVIDIA  RTX 6000 48\,GB GPU. The static and quasi-static 2D networks each required approximately 2.5\,h and 4.5\,h for $2\times10^6$ iterations on 1000 training samples. The dynamic networks required approximately 6\,h per geometry. \WW{The 3D static case required approximately 4.5 h for $1 \times 10 ^ {6}$ epochs.} Training is a one-time offline cost; once pretrained, each NO evaluation costs $\mathcal{O}(10^{-2})\,\mathrm{s}$ per time step, independent of mesh refinement in $\Omega_{\mathrm{II}}$.

\section{Non-dimensionalized Equations}
\label{APP_II}

Let $\bar{x} = x/L_{\mathrm{ref}}$, $\bar{u} = u/U_{\mathrm{ref}}$, $\bar{\sigma} = \sigma/E_{\mathrm{ref}}$, and $\bar{t} = t\sqrt{E_{\mathrm{ref}}/\rho_{\mathrm{ref}}}/L_{\mathrm{ref}}$, where reference scales are $L_{\mathrm{ref}} = 1\,\mathrm{m}$, $E_{\mathrm{ref}} = 1000\,\mathrm{Pa}$, and $\rho_{\mathrm{ref}} = 5\,\mathrm{kg\,m^{-3}}$.
By substituting Eqs. \ref{eq:strain_linear} and \ref{eq:stress_linear} into Eq. \ref{eq:static}, static equilibrium equation for a linear elastic material in the absence of external forces is obtained as 
\begin{equation}
    \frac{(\lambda + \mu)}{E_{el} L} \nabla^*(\nabla^* \cdot \mathbf{u}^*) + \frac{\mu}{E_{el} L} {\nabla^*}^2 \mathbf{u}^* = \mathbf{0} \nonumber
\end{equation}
where the superscript $^*$ indicates a non-dimensional quantity or operator; $L =1~\text{cm}$  is the reference length, and $E_{el} 2 \times 10^6~\text{Pa} $ is the reference elastic Young's modulus.\\
Correspondingly, the non-dimensional dynamic equilibrium equation takes the form 
\begin{equation}
    \frac{(\lambda + \mu)}{E_{el} L} \nabla^*(\nabla^* \cdot \mathbf{u}^*) + \frac{\mu}{E_{el} L} {\nabla^*}^2 \mathbf{u}^* = \frac{L^3 \rho}{m_0 t_0^2} \ddot{\mathbf{u}}^* \nonumber
\end{equation}
with $m_0 = 1~\text{g}$ and $t_0 = 4 \times 10^{-4}~\text{s}$ defined as the reference mass and reference time, respectively.\\
Under quasi-static loading at time step $n$ and in the absence of external forces, the non-dimensional counterpart of Eq. \ref{eq:qs_equil} is expressed as:
\begin{equation}
    \frac{1}{L} \nabla^* \cdot \left[ \frac{\mu}{E_{hp}} (\mathbf{I} + \nabla^* \mathbf{u}^{n*}) + \frac{\left( \lambda \log J - \mu \right)}{E_{hp} J} \text{Cof}(\mathbf{I} + \nabla^* \mathbf{u}^{n*}) \right] = \mathbf{0} \nonumber
\end{equation}
where $\text{Cof}(\cdot)$ denotes the cofactor matrix and $E_{hp} = 1~\text{Pa}$ is the reference hyperelastic Young's modulus.

\end{document}